%% file: thesis_final.tex
\documentclass[11pt,notitlepage]{report}
\usepackage{bibunits}
\usepackage{hyperref}
\usepackage{uwmthesis}
\usepackage{graphicx}
\usepackage{color}
\usepackage{amsmath}
\usepackage{amssymb}
\usepackage{amsfonts}
\usepackage{rotating}
\usepackage{tensor}
\usepackage{ulem}

\usepackage{bm}
\usepackage{epsfig}
\usepackage{latexsym}

\newcommand{\postscript}[2]{\setlength{\epsfxsize}{#2\hsize}
   \centerline{\epsfbox{#1}}}

\newcommand{\lwig}{\mbox{\,\raisebox{.3ex}
    {$<$}$\!\!\!\!\!$\raisebox{-.9ex}{$\sim$}\,}}

\def\k{\kappa}                    

\def\m{\mu}
\def\n{\nu}
  
\def\6{\partial}

\def\thalf{\tfrac{1}{2}}

\def\av#1{\langle #1\rangle}

\newcommand{\be}{\begin{equation}}
\newcommand{\ee}{\end{equation}}
\newcommand{\beq}{\begin{equation}}
\newcommand{\eeq}{\end{equation}}
\newcommand{\bea}{\begin{eqnarray}}
\newcommand{\eea}{\end{eqnarray}}

\begin{document}
\title{
\bf \LARGE Manifestations of String Theory\\ in astrophysical data and at the LHC}
\author{\bf \Large Satoshi Nawata}
\majorprof{Luis  Anchordoqui}
\submitdate{August 2008}
\degree{Doctor of Philosophy}
\program{Physics}
\copyrightyear{2008}
\majordept{Physics}
\havededicationfalse
\haveminorfalse
\copyrightfalse
\doctoratetrue
\figurespagetrue
\tablespagefalse

\Abstract{

  With the advent of the LHC and the
  continuing influx of cosmological data, phenomenological aspects of
  string theory have received renewed attention in recent years and
  many problems have been properly incorporated in this framework. For
  instance, recent theoretical considerations in string theory have
  applied a statistical approach to the enormous landscape of metastable
  vacua. The large number of vacua may shed some light on the
  cosmological constant problem. In addition, in string theory,
  attempts have been made to address the hierarchy problem within the
  context of the existence of large or warped internal dimensions
  transverse to a braneworld where we are confined, which lowers the
  effective scale of gravity to the TeV region. If this were the case,
  unseen dimensions of the space-time can be at the border of the
  energy domain within reach of the next generation of particle
  accelerators.

  Although the picture of the landscape may be the key to the
  cosmological constant problem, it is well-known that the
  compactification of a string background to a four dimensional
  solution undergoing accelerating expansion is difficult, which is
  described by the no-go theorem of Maldacena-Nu\~nez.  In the first
  part of this Dissertation, we investigate the cosmological content of the
  Salam-Sezgin supergravity which circumvents one of the hypotheses of
  the no-go theorem of Maldacena-Nu\~nez and consequently can support
  a de Sitter phase when lifted to string theory. We find a
  solution to the field equations in qualitative agreement with the
  observed dark energy density. The carrier of the acceleration in the
  present de Sitter epoch is a quintessence field slowly rolling down
  its exponential potential. Intrinsic to this model, there is a
  second modulus, which is automatically stabilized and acts as a
  source of cold dark matter with a mass proportional to an
  exponential function of the quintessence field.

  In the second part, we explore a "new physics'' signal at LHC, in
  the processes $pp \rightarrow \gamma + {\rm jet}$, $pp \rightarrow
  \gamma \gamma$, and $pp \to {\rm dijet}$. In D-brane quivers, there
  are one or more additional $U(1)$ gauge symmetries, beyond the
  $U(1)_Y$ of the standard model, which follows from the property that
  the gauge group for open strings terminating on a stack of $N$
  identical D-branes is $U(N)$ rather than $SU(N)$ for $N>2.$ Because
  of this, the photon will participate with the $SU(N)$ gauge boson in
  string tree level scattering processes which in the standard model
  occur only at one-loop level. In order to evaluate this stringy
  correction, we considered the processes $gg\rightarrow g\gamma$ and
  $gg\rightarrow \gamma \gamma$, and found that cross section
  measurements of the process $pp \rightarrow \gamma + {\rm jet}$ at
  the LHC will attain $5\sigma$ discovery reach on low scale string
  models for $M_{\rm string}$ as large as 4~TeV.  We have also
  considered the processes $gg \to gg$, $gg \to q \bar q$, $q \bar q
  \to gg$, $q g \to g q$ and $\bar q g \to \bar q g$, and found that,
  for the $pp \to {\rm dijet}$ channel, the LHC discovery reach will
  extend up to $M_{\rm string} \sim 6.8$~TeV.}

\beforepreface \prefacesection{Preface} 

This Dissertation is based on four papers that explore the rich
interdisciplinary boundaries of string theory, particle physics, and
cosmology. 

\vspace*{0.5cm}

\noindent The cosmological set-up presented in Chapter 2 is based on: 

\vspace*{0.25cm}

\noindent  L.~Anchordoqui, H.~Goldberg, S.~Nawata and C.~Nu\~nez,
  {\em Cosmology from String Theory,}
  Phys.\ Rev.\  D {\bf 76}, 126005 (2007).

\vspace*{0.5cm}

\noindent Chapters 3 and 4 are based on material from:

\vspace*{0.25cm}

\noindent   L.~A.~Anchordoqui, H.~Goldberg, S.~Nawata and T.~R.~Taylor,
  {\em Jet signals for low mass strings at the Large Hadron Collider,}
Phys. Rev. Lett. {\bf 100}, 171603 (2008). 
  
\vspace*{0.25cm}

\noindent and

\vspace*{0.25cm}

\noindent   L.~A.~Anchordoqui, H.~Goldberg, S.~Nawata and T.~R.~Taylor,
{\em Direct photons as probes of low mass strings at the CERN LHC}, 
Phys. Rev. D {\bf 78}, 016005 (2008).  

\vspace*{0.5cm}

\noindent The ideas discussed in Chapter 5 are based on:

\vspace*{0.25cm}

\noindent L.~A.~Anchordoqui, H.~Goldberg, D.~L\"ust, S.~Nawata, S.~Stieberger, and T.~R.~Taylor, {\em Dijet signals of low mass strings at the LHC}, arXiv:0808.0497 [hep-ph].

\prefacesection{Acknowledgments}
\input{acknowledge}

\afterpreface

\Chapter{Introduction}
\label{ch:introduction}
\input{introduction}

\Chapter{String Cosmology}
\label{ch:sancle}
\input{chap_sancle}

\Chapter{Photon Signals of Low Mass Strings at the LHC}
\label{ch:macho}
\input{chap_jets}

\Chapter{Strings {\em vs.} Black Holes at the LHC}
\label{ch:findchirp}
\input{chap_bh}

\Chapter{Dijet Signals of Low Mass Strings at the LHC}
\input{chap_dijets}

\Chapter{Conclusion}
\label{ch:conclusion}
\input{conclusion}

\clearpage

\appendix




\Chapter{Proof of the no-go theorem }

For completness, in this Appendix, we provide the proof of the no-go
theorem of Maldacena and Nu\~nez~\cite{Maldacena:2000mw}.\\

\underline{Theorem of Maldacena and Nu\~nez}:\\
Consider a
$D$ dimensional gravity theory which is compactified on $d$
dimensions. If the $D$ dimensional gravity theory
satisfies the following conditions:
\begin{enumerate}
\item the gravity action does not contain higher curvature corrections;
\item the potential is non-positive, $V \leq 0$;
\item the theory contains massless fields with positive kinetic terms; 
\item the internal manifold is compact without boundary and its volume is 
finite;
\end{enumerate}
then there are no non-sigular spontaneous compactifications to Minkowwski or
de-Sitter of the form  \be ds^2=\Omega^2(y)\, (dx_d^2+\hat g_{mn}
dy^mdy^n) \, , \ee where $dx_d^2=\eta_{\m\n}dx^\m dx^\n$, $\eta$
is the metric of the $d$ dimensional space that is either Minkowski or
dS space, and $g_{mn} dy^mdy^n$ is the metric of the internal
space. (In this Appendix we denote by $L, M, N, \ldots$ the $D$
dimensional indices, by $\m, \n, \rho, \ldots$ the $d$-dimensional
ones, and by $l, m, n, \ldots$ the ones for the internal space.)\\

\underline{Proof}:\\
From assumption 1, we can write
Einstein's equations in $D$ dimensions \be
R_{MN}=T_{MN}-\frac{1}{D-2}\, g_{MN} \, T^{L}_{L} \ . \ee  
Using the metric form, 
Einstein's equations can be re-written as
\begin{eqnarray}
R_{\m\n} & = & R_{\m\n}(\eta) -\eta_{\m\n}\left(\hat \nabla^2\log 
\Omega + (D-2)(\hat \nabla \log \Omega)^2\right) \nonumber \\
& = & T_{\m\n}-\frac{1}{D-2}\Omega^2 \eta_{\m\n} T^{L}_{L}  \, ,
\end{eqnarray}
where $\hat\nabla$ is the covariant derivative operator of the metric ${\hat g}$, and its indices are contracted with $\hat g$. Taking the trace over $\eta$ on
both sides we find 
\begin{eqnarray} 
\hat \nabla^2\log \Omega+(D-2)(\hat
\nabla \log \Omega)^2 & = & \frac{1}{(D-2)\Omega^{D-2}}\hat \nabla
^2\Omega^{D-2} \nonumber \\ 
 & = & R(\eta)+ \Omega^2
\left(- T^\m_\m+\frac{d}{D-2}T^{L}_{L} \right) \ . 
\label{b} 
\end{eqnarray}
Now, we define 
\be
\tilde T \equiv -T^\m_\m+\frac{d}{D-2}T^{L}_{L} \, .
\ee
First, we shall show that $\tilde T$ is non-negative. For a given
potential of matter fields $V$, we have $T_{MN} \sim -V g_{MN}$, and
\be 
\tilde T \sim Vd -\frac{d}{D-2} DV=-\frac{2d}{D-2} V\ge0 \, , 
\ee 
where
the last equality follows from assumption 2.  The energy 
momentum tensor of $n$-form fields takes the form 
\be
T_{MN}=F_{ML_1\cdots L_{n-1}}F_N^{L_1\cdots L_{n-1}}
-\frac{1}{2n}g_{MN}F^2 \, ,
\ee 
which in turn gives 
\begin{equation}
T_\m^\m=F_{\m L_1\cdots
  L_{n-1}}F^{\m L_1\cdots L_{n-1}} -\frac{d}{2n} F^2 \, .  
\label{trace}
\end{equation}
Hence,
\be \tilde T=-F_{\m L_1\cdots L_{n-1}}F^{\m L_1\cdots L_{n-1}}
+\frac{d}{D-2}\left(1-\frac{1}{n} \right) F^2 \;. \label{a} \ee 
The space time indices of non-vanishing components of $F$ could be completely along the internal dimensions or, if $n \geq d$, they could have $d$ out of $n$ indices along the $d$ dimensions and the rest along the internal dimensions. Otherwise the isometry
of $R^d$ or $dS^d$ is broken. They separately contribute to $\tilde T$.
In the former case, $F^2 \geq 0$. It follows from Eq.~(\ref{trace}) that
$\tilde T \geq 0$ for $n>1$ and $\tilde T =0$ for $n=1$.
In the latter case, $F^2<0$ and 
\begin{equation}
F_{\mu L_1 \cdots L_{n-1}}
F^{\mu L_1 \cdots L_{n-1}} =\frac{d}{n} F^2 \, .
\end{equation} 
It again follows
\begin{equation}
\tilde T
= \Big[ -\frac{d}{n}+\frac{d}{D-2} \Big(1-\frac{1}{n} \Big)\Big] F^2 
= -\frac{d(D-2-n+1)}{n(D-2)}F^2 \geq 0.
\end{equation}
Consequently we have in general
\begin{equation}
\tilde T \geq 0.
\end{equation}
Combined with Eq.~(\ref{b}) and the assumption that our $d$-dimensional
space is Minkowski or de Sitter with non-negative scalar curvature,
this means that
\begin{equation}
\Omega^{D-2}\hat\nabla^2\Omega^{D-2} \geq 0.
\end{equation}
The equality holds only if the right hand side of Eq.~(\ref{b}) is zero so that the $d$ dimensional space is Minsowski space. Since the internal space is compact, $\Omega$ is bounded below and above. Hence, integrating this
over the internal space by parts, we obtain
\begin{equation}
\int d^{D-d}y \sqrt{-\hat g} (\hat\nabla \Omega^{D-2})^2 \leq 0 \, .
\end{equation}
The left hand side is positive-definite so that this is valid only if
$\Omega$ is constant and the equality holds. This implies that the
right hand side of Eq.~(\ref{b}) vanishes, hence dS space is not
allowed and the only $n$ forms that can be turned on are the $n=1$,
$D-1$ forms. On the other hand, since the spontaneous compactification
does not allow $\Omega={\rm constant}$, Minkowski space is also
forbidden. In addition, it follows that the effective Newton constant
is finite since the $d$ dimensional Newton constant is given by \be
\frac{1}{G^d_N} \sim \int d^dy \sqrt{\hat g}\Omega ^{(d-2)} \, . \ee

\clearpage

\Chapter{The String Connection}

In this Appendix we briefly comment on how the six dimensional
solution derived in Chapter 2 reads in string theory. To this end, we
use the uplifting formulae developed by Cvetic, Gibbons and
Pope~\cite{Cvetic:2003xr}; we will denote with the subscript ``cgp''
the quantities of that paper and with ``us,'' quantities in this
Dissertation.  Let us more specifically look at Eq.~(34) in
Ref.~\cite{Cvetic:2003xr}, where the authors described the six
dimensional Lagrangian they uplifted to Type I string theory. By
simple inspection, we can see that the relation between their
variables and fields with the ones we used in Eq.~(\ref{action}) is $
\phi|_{\rm cgp}= -2\phi|_{\rm us},$ $F_2|_{\rm cgp} = \sqrt{G_6}
F_2|_{\rm us},$ $H_3|_{\rm cgp} =\sqrt{G_6/3}\, G_{3}|_{\rm us},$ and
$\bar{g}^2|_{\rm cgp} = \xi/(8 G_6)|_{\rm us}.$ Our six dimensional
background is determined by the (string frame) metric $ ds_6^2 =
e^{2f}\, \Big[- dt^2 + e^{2h}dx_3^2 + r_c^2 \, d\sigma_2{}^2 \Big],$
the gauge field $F_{\vartheta \varphi}=-b \sin \vartheta,$ and the
$t$-dependent functions $h(t),$ $f(t)= \sqrt{G_4}\, (X-Y)/4,$ and
$\phi(t)=\sqrt{G_4}\, (X+Y)/2.$ Identifying these expressions with
those in Eqs.~(47), (48) and (49) of Ref.~\cite{Cvetic:2003xr}, one
obtains a full Type I or Type IIB configuration, consisting of a
3-form (denoted by $F_3$),
\begin{eqnarray}
F_3 & = & \frac{8 G_6 \sinh\hat \rho \cosh\hat \rho}{\xi \cosh^2 2\hat \rho}d\hat \rho \wedge
\Big(d\alpha -\sqrt{\frac{\xi}{8 G_6}} b
\cos\vartheta d\varphi \Big) \wedge \Big(d\beta +\sqrt{\frac{\xi}{8 G_6}} b
\cos\vartheta d\varphi \Big) \nonumber \\
& - & \frac{\sqrt{2} G_6 b}{\sqrt{\xi} \cosh 2\hat \rho} \sin\vartheta d\vartheta
\wedge
d\varphi \wedge \left[ \cosh^2\hat\rho \left(d\alpha -\sqrt{\frac{\xi}{8 G_6}} b
\cos\vartheta d\varphi \right) \right. \nonumber \\
& - & \left. \sinh^2\hat \rho \left(d\beta
+\sqrt{\frac{\xi}{8 G_6}} b
\cos\vartheta d\varphi \right)  \right] \,,
\end{eqnarray}
a  dilaton (denoted by $\hat{\phi}$)
\begin{equation}
e^{2\hat{\phi}}=\frac{e^{2\phi}}{\cosh(2\hat \rho)} \,,
\end{equation}
and a ten dimensional metric that in the
string frame
reads
\begin{eqnarray}
ds^2_{\rm str} &=& e^{\phi}\, ds_6^2 + dz^2 +\frac{4G_6}{\xi} \left[ d\hat \rho^2
+\frac{\cosh^2\hat \rho}{\cosh 2\hat \rho} \left( d\alpha -\sqrt{\frac{\xi}{8 G_6}} b
\cos\vartheta d\varphi \right)^2 \right. \nonumber \\
 & + & \left. \frac{\sinh^2\hat \rho}{\cosh 2\hat \rho}
\left(d\beta +\sqrt{\frac{\xi}{8 G_6}} b
\cos\vartheta d\varphi \right)^2\right] \,,
\label{config10d}
\end{eqnarray}
where $\hat \rho,\, z,\, \alpha,$ and $\beta$ denote the four extra
coordinates.  It is important to stress that though the uplifted
procedure decribed above implies a non-compact internal manifold, the
metric in Eq.~(\ref{config10d}) can be interpreted within the context
of~\cite{Giddings:2001yu} (i.e., $0 \leq \hat \rho \leq L,$ with
$L \gg 1$ an infrared cutoff where the spacetime smoothly closes up)
to obtain a finite volume for the internal space and consequently a
non-zero but tiny value for $G_6.$ 

\clearpage

\Chapter{Pole residues of the Veneziano form factor}

Consider the product of Gamma functions
\begin{equation}
\Gamma(n) \, \, \Gamma(1-n) = \frac{\pi}{\sin (n\pi)} \ \  .
\end{equation}
In the limit  $1-n = \epsilon \ll 1$, 
$\sin (n \pi) = \sin (\pi - \pi \epsilon) = 
\sin (\pi) - \pi \epsilon \cos (\pi) 
= \pi \epsilon$, and so
\begin{equation}
\Gamma ( 1 - \epsilon) \,  \, \Gamma (\epsilon) = \frac{\pi}{\pi \epsilon} = \frac{1}{\epsilon} \, ,
\end{equation}
which in turn leads to
\begin{equation}
\lim_{n \to 1} \Gamma (1 - n) = \frac{1}{1-n} \, .
\end{equation} 
Therefore, in the limit of $s \to 1$,
\begin{equation}
\mu (s, t, u) = \frac{\Gamma (1 -u)}{\Gamma (1 + t)} \frac{1}{(1-s)} = 
\frac{\Gamma (2+t)}{\Gamma (1 + t)} \, \frac{1}{1-s} = \frac{1+t}{1+s} 
\label{label1}
\end{equation}
and
\begin{equation}
\mu (s,u,t) =  \frac{\Gamma (1 -t)}{\Gamma(1 + u)} 
\frac{1}{(1-s)} =
\frac{\Gamma (2+u)}{\Gamma (1 + u)} \,\frac{1}{1-s}
= \frac{1+u}{1 +s} \, .
\label{label2}
\end{equation}
We can now expand the string squared amplitude,
\begin{eqnarray}
|{\cal M} (gg \to \gamma g)|^2 & = & 
\left| \frac{s}{u} \, \mu(s,t,u) + \frac{s}{t} \, \mu(s,u,t) \right|^2 
+ \left| \frac{t}{u} \, \mu(s,t,u) + \frac{t}{s} \, \mu(t,u,s) \right|^2 
\nonumber \\
& + & \left|\frac{u}{s} \, \mu(u,t,s) + \frac{u}{t} \, \mu(u,s,t) 
\right|^2 \,,
\end{eqnarray}
near the pole yielding
\begin{eqnarray}
|{\cal M} (gg \to \gamma g)|^2 & \propto & \left| \frac{s_0}{u} \, \, \frac{\Gamma(1-u)}{\Gamma(1+t)} + \frac{s_0}{t}  \, \, 
\frac{\Gamma(1-t)}{\Gamma(1+u)} \right|^2 \,
\frac{1}{(s - s_0)^2} \nonumber \\
& + & \left| -\frac{t}{u} \, \, \frac{\Gamma(1-u)}{\Gamma(1+t)} + 
\frac{t}{s_0} \, \, \left[ \frac{\Gamma(1-t)}{\Gamma(1+u)} - \frac{\Gamma(1-u)}{\Gamma(1+t)}\right] \right|^2 \, \, \frac{1}{(s - s_0)^2} \nonumber \\
   & + &  \left| \frac{u}{s_0} \, \, \left[ 
\frac{\Gamma(1-u)}{\Gamma(1+t)} -\frac{\Gamma(1-t)}{\Gamma(1+u)}\right] - \frac{u}{t}  \, \,  \frac{\Gamma(1-t)}{\Gamma(1+u)} \right|^2 \, \,
\frac{1}{(s - s_0)^2} \, \, ,
\end{eqnarray}
where we have restored the string scale, $s_0 = M_s^2.$ 
Equivalently,
\begin{eqnarray}
|{\cal M} (gg \to \gamma g)|^2 & \propto & \left\{ \left| \frac{s_0}{u} \ A + \frac{s_0}{t} \ B \right|^2 +\left|-\frac{t}{u}  \ A + \frac{t}{s_0} \ (B-A) \right|^2 \right. \nonumber \\ 
 & + & \left. \left|\frac{u}{s_0} \ (A-B) - \frac{u}{t} \ B \right|^2 \right\} 
\ \ \frac{1}{(s - s_0)^2} \, \,,
\label{ponyU}
\end{eqnarray}
where
\begin{equation}
A = \frac{\Gamma(1-u)}{\Gamma (1+t)} = 1 +t = -u
\end{equation}
and
\begin{equation}
B = \frac{\Gamma(1-t)}{\Gamma(1+u)} = 1 + u = -t 
\end{equation}
are obtained from Eqs.~(\ref{label1}) and (\ref{label2}).
Then, Eq.~(\ref{ponyU}) becomes
\begin{eqnarray}
|{\cal M} (gg \to \gamma g)|^2 & \propto & 4 s_0^2 +  \left| t + \frac{t}{s_0} (-t + u)\right|^2 + \left| u + \frac{u}{s_0} (-u + t)\right|^2 \nonumber \\
 & \propto &  \left[4 s_0^2 + \frac{4t^4 + 4u^4}{s_0^2}\right]  \,\, \frac{1}{(s-s_0)^2} \, \,,
\end{eqnarray}
where we have used the Mandelstam relation: $u = -s_0 -t$.  Finally,
the singularity is smeared with a width $\Gamma$ to obtain
Eq.~(\ref{mhvlow2}).

\clearpage

\Chapter{Invariant mass spectrum}
In this Appendix D, we shall derive the invariant mass formula Eq. (\ref{longBH}). For this purpose, we write the total cross-section for the process $pp\rightarrow \gamma + {\rm jet}$:
\begin{equation}\label{totalcs}
\left. \sigma \right|_{pp\rightarrow \gamma+ {\rm jet}} =\int^1_0dx_a \int^1_0dx_b \sum_{ijk} f_i(x_a) f_j(x_b) \left. \sigma \right|_{ij\rightarrow k,\gamma}
\end{equation}
Relation (\ref{longitudinal}) lets us convert the integral in Eq. (\ref{totalcs}) into an integral over the parameters $M^2$, $Y$. The Jacobian of the change of variables is
\begin{eqnarray}
\frac{\partial(M^2, Y)}{\partial(x_a,x_b)}=\left|\begin{array}{ccc} x_bs & x_bs \\
\thalf x_a & -\thalf x_b \\ \end{array} \right| =s \, .
\end{eqnarray}
Hence, we obtain
\begin{equation}
\left. \frac{d\sigma}{dM^2} \right|_{pp\rightarrow \gamma+ {\rm jet}} =\frac{1}{s} \int \! dY \! \int d\hat t \;  f_i(x_a) f_j(x_b) \left. \frac{d\sigma}{d\hat t} \right|_{ij\rightarrow k,\gamma}  \, .
\end{equation}
Eq. (\ref{Mandelstam}) gives us 
\begin{equation}
\frac{d\hat t}{dy}= \frac{M^2}{2} \frac{1}{\cosh^2 y} \, ,
\end{equation}
so that the invariant mass spectrum can be written as
\begin{equation}
\left. \frac{d\sigma}{dM} \right|_{pp\rightarrow \gamma+ {\rm jet}} =\frac{M^3}{s} \int \! dY \! \int \frac{dy}{\cosh^2 y}\;  f_i(x_a) f_j(x_b) \left. \frac{d\sigma}{d\hat t} \right|_{ij\rightarrow k,\gamma} \, .
\label{invmass}
\end{equation}
Since we set cuts in the photon and jet rapidities
\begin{equation}
|y_1| = |y+Y| < y_{\rm max} = 2.4, \ \ \ \ \ \ |y_2| 
= |y-Y| < y_{\rm max} = 2.4 \, ,
\end{equation}
Eq. (\ref{invmass}) can be expressed with integral limits as Eq. (\ref{longBH}).


\clearpage
\birthplacedate{Tokyo, Japan \>\>February 8, 1982}
\collegewherewhen{%
\> Tokyo Institute of Technology \>\> 2002-2004, \>BA\\
\>\uwm	\>\>2005-2008, \>Ph.D.}

\newpage

\addcontentsline{toc}{chapter}{\numberline {Curriculum Vitae}}
\null\vskip1in%
\begin{center}
{\Large\bf Curriculum Vitae}
\end{center}
\vskip 2em
\begin{tabbing}
\tabset
Title of Dissertation\\
\>{\bf \em Manifestations 
of String Theory in Astrophysical Data and at the LHC}
\end{tabbing}
\vskip 1em

\begin{startvita}
\end{startvita}

\renewenvironment{thebibliography}[1]%
  {\begin{list}{\labelenumi\hss}%
     {\usecounter{enumi}\setlength{\labelwidth}{3em}%
      \setlength{\leftmargin}{5em}}}%
  {\end{list}}
\renewcommand{\bibitem}[1]{\item\label{#1}\relax}%
\renewcommand{\theenumi}{\arabic{enumi}}%
\begin{publications}
\begin{itemize}
\item L.~Anchordoqui, H.~Goldberg, S.~Nawata and C.~Nu\~nez,
  {\em Cosmology from String Theory,}
  Phys.\ Rev.\  D {\bf 76}, 126005 (2007)
  [arXiv:0704.0928 [hep-ph]].
\item L.~A.~Anchordoqui, H.~Goldberg, S.~Nawata and T.~R.~Taylor,
  {\em Jet signals for low mass strings at the Large Hadron Collider,}
  Phys. Rev. Lett. {\bf 100}, 171603 (2008) 
  [arXiv:0712.0386 [hep-ph]].
\item L.~A.~Anchordoqui, H.~Goldberg, S.~Nawata and T.~R.~Taylor,
{\em Direct photons as probes of low mass strings at the CERN LHC,}
 Phys. Rev D {\bf 78}, 016005 (2008) 
  [arXiv:0804.2013 [hep-ph]].
\item L.~A.~Anchordoqui, H.~Goldberg, D.~L\"ust, S.~Nawata,
  S.~Stieberger, and T.~R.~Taylor, {\em Dijet signals of low mass
    strings at the LHC}, arXiv:0808.0497 [hep-ph].

\end{itemize}
\end{publications}
\begin{honorarysocieties}
2008\>
Institute for Advanced Study, School of Natural Sciences, Princeton, NJ\\
\>Attendance Fellowship: ``Strings and Phenomenology''\\
2008\>Kavli Institute for Theoretical Physics, University of California, 
Santa Barbara, CA\\
\> Attendance Fellowship: ``Gauge Theory and Langlands Duality''\\
2008\> UWM Chancellor's Graduate Student Fellowship\\
2007 \> Institute for Advanced Study, School of Mathematics, Princeton, NJ\\
\>Attendance Fellowship: ``Workshop on Gauge Theory and Representation Theory''\\ 
2007\> UWM Chancellor's Graduate Student Fellowship\\
2006\> UWM Chancellor's Graduate Student Fellowship\\
2005 \> UWM Chancellor's Graduate Student Fellowship\\
\end{honorarysocieties}

\finishvita
\end{document}

%% file: acknowledge.tex

First and foremost, I would like to express my deep gratitude to my
adviser, Professor Luis Anchordoqui.  His patience, educational
guidance, time and knowledge were paramount to my work within the last
2 years and this Dissertation would certainly not have been possible
without him.  Further, his willingness to introduce me to so many
distinguished colleagues in the field of particle physics, notably
Haim Goldberg, Dieter L\"ust, Carlos Nu\~nez, Stephan Stieberger and
Tomasz Taylor is also very much appreciated.  The papers we have
collaborated on have already allowed me to hone my research and
analytical skills, thereby opening other avenues to future research
and projects.

I am also thankful for the education and guidance from Distinguished
Professor John Friedman and Distinguished Professor Leonard Parker, as
well as the opportunity to work with everyone in the Center for
Gravitation and Cosmology.

I have also benefited greatly from the time and guidance provided to
me by Professor Richard Sorbello, in his role as the Graduate Program
Advisor, and from Ms. Kate Valerius, Graduate Program Assistant, whose
level of patience and willingness to help in so many of the
non-academic areas (keeping me organized, teaching communication and
writing skills, etc.) has truly been appreciated and will be
remembered for a long time to come.

%% file: introduction.tex
Elementary Particle Physics seeks to understand, at the deepest level,
the structure of matter and the forces by which it interacts. The
experimental success of the standard model (SM) of weak,
electromagnetic, and strong interactions can be considered as the
triumph of the gauge symmetry principle to describe all physical
phenomena up to energies $\sim 500$~GeV~\cite{Yao:2006px}. However,
the SM remains unsatisfactory in some of its theoretical aspects. The
major one concerns quantum gravity effects: the
renormalization procedure that allows one to extract finite
predictions for processes involving the three other fundamental forces
fails when gravitational interactions are taken into account. String
theory stands here as the only known consistent framework to
incorporate these effects, replacing the elementary point particles
(which form matter and mediate interactions) with a single extended
object of vanishing width~\cite{Green:1987sp,Green:1987mn}.  The known
fundamental particles appear ``point-like'' because the experimental
energies probed thus far by colliders are too small to excite the
string oscillation modes, so only the center of mass motion is
perceived. In addition to these heavy oscillation modes, strings have
new degrees of freedom that often take the classical geometry
description of propagation into ``hidden'' compact dimensions,
recovering the old ideas of Kaluza and
Klein~\cite{Kaluza:1921tu,Klein:1926tv}.

The distance at which quantum gravity comes into play is
unknown. Though this distance can be very small, ${\cal O}
(10^{-35}~{\rm m})$, a particularly interesting possibility arises if
it is ${\cal O} (10^{-19}~{\rm m})$, the distance at which the
electromagnetic and weak forces are known to unify to form the
electroweak force.  Lowering the scale of quantum gravity into the TeV
region provides a framework for solving the mass hierarchy problem and
unifying all interactions. The apparent weakness of gravity can be
accounted for by the existence of large~\cite{Arkani-Hamed:1998rs} or
warped~\cite{Randall:1999ee} internal dimensions transverse to a
braneworld where we are confined. If this were the case, spacetime's
unseen dimensions could be at the border of the energy domain within
reach of the next generation of particle accelerators, begining this
year with the Large Hadron Collider (LHC) at CERN. In particular, the
mass scale $M_s$ of fundamental strings would be as low as few
TeV~\cite{Antoniadis:1998ig}. This mass determines the center of mass
energy threshold $\sqrt{\hat s}\ge M_s$ for the production of Regge
resonances in parton collisions, thus for the onset of string effects
at the LHC~\cite{stringhunter}.  In this Dissertation we consider the
extensions of the SM based on open strings ending on D-branes, with
gauge bosons due to strings attached to stacks of D-branes and chiral
matter due to strings stretching between intersecting
D-branes~\cite{reviews}.  Only one assumption is necessary in order to
set up a solid framework: the string coupling must be small in order
to rely on perturbation theory in the computations of scattering
amplitudes. In this case, black hole production and other strong
gravity effects occur at energies above the string scale; therefore at
least few lowest Regge recurrences are available for examination, free
from interference with some complex quantum gravitational phenomena.
Starting from a small string coupling, the values of the SM coupling
constants are determined by D-brane configurations and the properties
of extra dimensions, hence that part of superstring theory requires
intricate model-building; however, as we show in this Dissertation,
some basic properties of Regge resonances like their production rates
and decay widths are completely model-independent.  The resonant
character of parton cross sections should be easy to detect at the LHC
if the string mass scale is not too high.

On a separate track, the remarkable accuracy of the Wilkinson
Microwave Anisotropy Probe (WMAP) five-year observations has
catapulted us into a new era in cosmology~\cite{Spergel:2006hy}. This
information, taken together with Big Bang nucleosynthesis (BBN)
abundances~\cite{Olive:1999ij}, astronomical observations charting the
large scale distribution of
galaxies~\cite{Cole:2005sx,Tegmark:2006az}, and luminosity distance
measurements of Type Ia
supernovae~\cite{Riess:1998cb,Perlmutter:1998np,Bahcall:1999xn}, seems
to ensure the existence of some unknown form of ``dark'' energy
density that dominates the recent gravitational dynamics
of the universe and yields a stage of cosmic acceleration. Moreover,
these impressive experiments have weighed the universe and determined
that the known particles make up only 5\% of its mass, providing
overwhelming evidence for new particles and fundamental laws of
nature.  The discovery of the two unknown components of the universe,
the so-called ``dark matter'' and dark energy, pose an important
challenge for particle physics that would be met in a few years by the
LHC~\cite{Peskin:2007nk}.  Dark matter appears to consist of
non-relativistic particles that only interact gravitationally and
perhaps by weak interactions. The nature of the second unknown
component is even less clear. The simplest candidate for such a
missing energy is a positive cosmological constant $\Lambda$. Such an
identification, however, unavoidably raises a series of questions:
$(a)$ Why is $\Lambda$ so small in particle physics units? -- the
so-called fine-tuning problem for $\Lambda$; $(b)$ Why is $\Lambda
\sim $ the present value (in Planck units) of the (dark) matter
density? -- the so-called ``coincidence problem.''  At present the most
promising scenarios for answering (at least part of) these questions
associate dark energy with a dynamical scalar
field~\cite{Ratra:1987rm}, generally called
``quintessence''~\cite{Zlatev:1998tr}, whose potential goes to zero
asymptotically (leaving therefore just the usual puzzle of why the
``true'' cosmological constant vanishes). The scalar field slowly
rolls down such a potential reaching infinity (and zero energy) only
after an infinite (or very long) time. While doing so quintessence
produces an effective, time-dependent, cosmic energy density
accompanied by a sufficiently negative pressure, i.e., an 
effective cosmological constant. As far as identifying quintessence is
concerned several possibilities have been considered (see
e.g.,~\cite{Ferreira:1997au,Ferreira:1997hj,Copeland:1997et,Peebles:1998qn,Parker:1999td,Parker:1999ac,Parker:2003as,Caldwell:2005xb});
in particular, those motivated by string theory include the dilaton
and time-varying moduli fields~\cite{Gasperini:2001pc}.  In addition
to analyzing for new signals at the LHC, this Dissertation is aimed
at investigating a number of topics that exploit the recent
astrophysical observations mentioned above. Our overall goal is of
furthering the understanding of cosmology while simultaneously
investigating new areas of fundamental physics.

The Dissertation is organized as follows. In the next chapter, we
study the cosmological content of Salam-Sezgin~\cite{Salam:1984cj} six
dimensional supergravity, which circumvents the no-go theorem of
Maldacena-Nu\~nez when the six dimensional space time is embedded in
Type I or Heterotic supergravity, in both cases with non-compact
extra-dimensions: ${\cal H}^{2,2}\times S^1$. We solve the field
equations matching the existing experimental data, and we find
candidates for the quintessence field and cold dark matter, which can
be written as linear combinations of the $S^2$ moduli field and the
six dimensional dilaton field.  In Chapters 3, 4, and 5 we discuss
phenomenological aspects of low mass string theory related to
experimental searched for physics beyond the SM at the LHC. Using a
generic property of D-brane quiver models, we search for Regge
recurrences at parton collision energies $\sqrt{\hat s} \sim M_s$. In
these models, the photon mixes with the hypercharge and the gauge
field of baryon number; hence some processes like $gg \to g \gamma$ or
$gg \to \gamma \gamma$ show up at the string disk level. In Chapter 3
we analyze $pp \to \gamma + {\rm jet}$ and $pp \to \gamma \gamma$
channels. By setting a high energy cut (300 GeV) on the transverse
momentum of the photon, we find a signal at the LHC which could probe
deviation from the SM at a $5\sigma$ significance for $M_s$ as large
as 2.3 TeV. After that, in Chapter 4 we study additional LHC
observables and discuss potential methods to discriminate string
resonances from other sources of new physics.  We first study the
invariant mass spectrum of string resonances in the $pp \to \gamma +
{\rm jet}$ channel and find that bump searches could help to increase
the LHC discovery reach up to $M_s \sim 4$~TeV. Then, we compare
$\gamma$ and $Z$ production via Hawking evaporation of TeV-scale back
holes and string excitations of D-brane models. In Chapter 5, we
extend our search of string signals at the LHC by analyzing the
process $pp \to {\rm dijet}$. We find that this channel exhibits a
resonant behavior at a $5\sigma$ significance for $M_s$ as large as
6.8~TeV. Chapter 6 contains the main conclusions of this Dissertation.
In Appendix A, we present the proof of the no-go theorem of
Maldacena-Nu\~nez.  In Appendix B we discuss the the connection
between Salam-Sezgin six dimensional supergravity and string
theory. In Appendices C and D, we collect calculations that are
somewhat technical for the main text.

%% file: chap_sancle.tex
The mechanism involved in generating a very small cosmological
constant that satisfies 't~Hooft naturalness is one of the most
pressing questions in contemporary physics. Recent observations of
distant Type Ia
supernovae~\cite{Riess:1998cb,Perlmutter:1998np,Bahcall:1999xn}
strongly indicate that the universe is expanding in an accelerating
phase, with an effective de-Sitter (dS) constant $H$ that nearly
saturates the upper bound given by the present-day value of the Hubble
constant, i.e., $H \leq H_0 \sim 10^{-33}$~eV. According to the
Einstein field equations, $H$ provides a measure of the scalar
curvature of the space and is related to the vacuum energy density
$\rho_{\rm vac}$ through Friedmann's equation, $3\, M_{\rm Pl}^2 H^2
\sim \rho_{\rm vac},$ where $M_{\rm Pl} \simeq 2.4 \times 10^{18}~{\rm
  GeV}$ is the reduced Planck mass.  However, the ``natural'' value of
$\rho_{\rm vac}$ coming from the zero-point energies of known
elementary particles is found to be at least $\rho_{\rm vac} \sim {\rm
  TeV}^4.$ Substitution of this value of $\rho_{\rm vac}$ into
Friedmann's equation yields $H \geq 10^{-3}$~eV, grossly inconsistent
with the set of supernova (SN) observations. The absence of a
mechanism in agreement with 't~Hooft naturalness criteria then centers
on the following question: why is the vacuum energy needed by the
Einstein field equations 120 orders of magnitude smaller than any
``natural'' cut-off scale in effective field theory of particle
interactions, but not zero?

Today, the most popular framework that can address aspects of this
question is the anthropic approach, in which the fundamental constants
are not determined through fundamental reasons, but rather because
such values are necessary for life (and hence intelligent observers to
measure the constants)~\cite{Weinberg:dv}. Of course, in order to
implement this idea in a concrete physical theory, it is necessary to
postulate a multiverse in which fundamental physical parameters can
take different values. Recent investigations in string theory have
applied a statistical approach to the enormous ``landscape'' of
metastable vacua present in the
theory~\cite{Bousso:2000xa,Susskind:2003kw,Douglas:2003um,Douglas:2006es}.
A vast ensemble of metastable vacua with a small positive effective
cosmological constant that can accommodate the low energy effective
field theory of the SM has been found. Therefore, the idea of a
string landscape has been used to propose a concrete implementation
of the anthropic principle.

Nevertheless, the compactification of a string/M-theory background to
a four dimensional solution undergoing accelerating expansion has
proved to be exceedingly difficult. The obstruction to finding dS
solutions in the low energy equations of string/M theory is well known
and summarized in the no-go theorem
of~\cite{Maldacena:2000mw,Gibbons:1984kp}.  This theorem states that
in a $D$-dimensional theory of gravity, in which $(a)$ the action is
linear in the Ricci scalar curvature, $(b)$ the potential for the
matter fields is non-positive, and $(c)$ the massless fields have
positive defined kinetic terms, there are no (dynamical)
compactifications of the form: $ds^2_D = \Omega^2(y) (dx_d^2 + \hat
g_{mn} dy^n dy^m)$, if the $d$ dimensional space has Minkowski
$SO(1,d-1)$ or dS $SO(1,d)$ isometries and its $d$ dimensional
gravitational constant is finite (i.e., the internal space has finite
volume). Further details are given in the Appendix~A.  The conclusions of the
theorem can be circumvented if some of its hypotheses are not
satisfied. Examples where the hypotheses can be relaxed exist: $(i)$
one can find solutions in which not all of the internal dimensions are
compact~\cite{Gibbons:2001wy}; $(ii)$ one may try to find a solution
breaking Minkowski or de Sitter invariance~\cite{Townsend:2003fx};
$(iii)$ one may try to add negative tension matter (e.g., in the form
of orientifold planes)~\cite{Giddings:2001yu}; $(iv)$ one can even
appeal to some intrincate string dynamics~\cite{Kachru:2003aw}.

The Salam-Sezgin six dimensional supergravity model~\cite{Salam:1984cj}
provides a specific example where the no-go theorem is not at work,
because when their model is lifted to M theory the internal space is
found to be non-compact~\cite{Cvetic:2003xr} (See Appendix B). The
lower dimensional perspective of this, is that in six dimensions the
potential can be positive.  This model has perhaps attracted the most
attention because of the wide range of its phenomenological
applications (see
e.g.,~\cite{Halliwell:1986bs,Aghababaie:2002be,Aghababaie:2003wz,Gibbons:2003di,Aghababaie:2003ar}). In
this chapter we examine the cosmological implications of such a
supergravity model during the epochs subsequent to primordial
nucleosynthesis. We derive a solution of Einstein field equations 
that is in qualitative agreement with luminosity distance
measurements of Type Ia
supernovae~\cite{Riess:1998cb,Perlmutter:1998np,Bahcall:1999xn},
primordial nucleosynthesis abundances~\cite{Olive:1999ij,Bean:2002sm},
data from the Sloan Digital Sky Survey (SDSS)~\cite{Tegmark:2003ud},
and the most recent measurements from the Wilkinson Microwave
Anisotropy Probe (WMAP) satellite~\cite{Spergel:2006hy}. The observed
acceleration of the universe is driven by the ``dark energy''
associated to a scalar field slowly rolling down its exponential
potential (i.e., kinetic energy density $<$ potential energy density
$\equiv$ negative pressure). Very interestingly, the resulting
cosmological model also predicts a cold dark matter (CDM) candidate.
In analogy with the phenomenological proposal
of~\cite{Comelli:2003cv,Franca:2003zg}, such a nonbaryonic matter
interacts with the dark energy field and therefore the mass of the CDM
particles evolves with the exponential dark energy potential.
However, an attempt to saturate the present CDM component in this
manner leads to gross deviations from present cosmological data. We
will show that this type of CDM can account for up to about 7\% of the
total CDM budget.  Generalizations of our scenario (using
supergravities with more fields) might account for the rest.

\section{Salam-Sezgin Cosmology}

We begin with the action of Salam-Sezgin six dimensional
supergravity~\cite{Salam:1984cj}, setting to zero the fermionic terms
in the background (of course fermionic excitations will arise from
fluctuations),
\beq S=\frac{1}{4 \kappa^2}\int d^6x \sqrt{g_6}\Big[ R
- \kappa^2 (\partial_M\sigma)^2 -\kappa^2 e^{\kappa\sigma}F_{MN}^2
-\frac{2g^2}{\kappa^2}e^{-\kappa\sigma}
-\frac{\kappa^2}{3}e^{2\kappa\sigma} G_{MNP}^2 \Big] \,\, .
\label{ss}
\eeq
Here, $g_6=\det g_{MN},$ $R$ is the Ricci scalar of $g_{MN},$
$F_{MN}=\partial_{[M} A_{N]},$ $G_{MNP}=\partial_{[M}B_{NP]}
+\k A_{[M} F_{NP]},$ and capital Latin indices run from 0 to 5.
A re-scaling of the constants:
$G_6\equiv2 \kappa^2,$ $\phi \equiv -\k\sigma$ and $\xi\equiv 4\,g^2$
leads to
\beq
S=\frac{1}{2 G_6} \int d^6x\sqrt{g_6}\Big[R -  (\partial_M\phi)^2 -
\frac{\xi}{G_6} e^\phi - \frac{G_6}{2} e^{-\phi} F_{MN}^2 -
\frac{G_6}{6}
e^{-2\phi} G_{MNP}^2\Big] \,\,.
\label{action}
\eeq
The length dimensions of the
fields are: $[G_6]=L^4,$ $[\xi]=L^2,$ $[\phi]=[g_{MN}^2]=1,$
$[A_M^2]=L^{-4},$ and $[F_{MN}^2]=[G_{MNP}^2]= L^{-6}.$

Now, we consider a
spontaneous compactification from six dimension to four dimension.
To this end, we take the six dimensional manifold $M$ to be
a direct product of 4 Minkowski directions (hereafter denoted by
$N_1$) and a compact orientable two dimensional manifold $N_2$ with
constant curvature.  Without loss of generality, we can set
$N_2$ to be a sphere $S^2$, or a $\Sigma_2$ hyperbolic manifold
with arbitrary genus. The metric on $M$ locally takes the form
\bea ds_6^2= ds_4(t,{\vec x})^2 + e^{2f(t,{\vec x})}d\sigma^2, &&
d\sigma^2=\left\{\begin{array}{ll}
   r_c^2\, (d\vartheta^2 +\sin^2\vartheta d\varphi^2) & \,\, {\rm for} \ S^2 \\
   r_c^2\, (d\vartheta^2 +\sinh^2\vartheta d\varphi^2) & \,\, {\rm for}\
\Sigma_2 \,\,,
  \end{array} \right.
\label{metric}
\eea
where $(t,\ \vec x)$ denotes a local coordinate system in $N_1,$ $r_c$ is
the compactification radius of $N_2$.
We assume that the scalar field $\phi$ is only
dependent on the point of $N_1$, i.e., $\phi=\phi(t,\ \vec x)$. We further
assume that the gauge field $A_M$ is excited on $N_2$ and is of the form
\bea
A_\varphi=\left\{\begin{array}{ll}
    b\cos \vartheta &(S^2) \\
    b\cosh \vartheta& (\Sigma_2) \, .
  \end{array} \right. \label{FC}
\eea
This is the monopole configuration detailed by
Salam-Sezgin~\cite{Salam:1984cj}. Since
we set the Kalb-Ramond field $B_{NP} = 0$ and the term
$A_{[M} F_{NP]}$ vanishes on $N_2$, $G_{MNP} = 0$.
The field strength becomes
\begin{equation}
F_{MN}^2=
    2b^2 e^{-4f}/r_c^4  \, .
\label{FS}
\end{equation}
Taking the variation of the gauge field $A_M$ in Eq.~(\ref{action})
we obtain the Maxwell equation
\be
\partial_M \Big[ \sqrt{g_4}\sqrt{g_\sigma} e^{2f-\phi} F^{MN}  \Big]=0.
\label{Maxwell}
\ee
It is easily seen that the field strengths in Eq.~(\ref{FS}) satisfy
Eq.~(\ref{Maxwell}).

With this in mind, the Ricci scalar reduces to~\cite{Wald:1984rg} \be
R[M]=R[N_1]+e^{-2f}R[N_2]-4\Box f-6(\partial_\m f)^2 \,\,, \ee where
$R[M],$ $R[N_1],$ and $ R[N_2]$ denote the Ricci scalars of the
manifolds $M,$ $N_1,$ and $N_2$; respectively. (Greek indices run from
0 to 3). The Ricci scalar of $N_2$ reads \bea R[N_2]=
\left\{\begin{array}{ll}
    +2/r_c^2 & (S^2)\\
    -2/r_c^2 & (\Sigma_2).
\end{array} \right.
\label{007}
\eea
To simplify the notation, from now on, $R_1$ and $R_2$ indicate $R[N_1]$
and $R[N_2]$, respectively. The determinant of the metric can be written as
$\sqrt{g_6}=e^{2f}\sqrt{g_4}\sqrt{g_\sigma},$
where $g_4=\det g_{\m\n}$ and $g_\sigma$ is the determinant of the metric
of $N_2$ excluding the factor $e^{2f}$.
We define the gravitational constant in the four dimension as
\be
\frac{1}{G_4}\equiv \frac{M_{\rm Pl}^2}{2} = \frac{1}{2\, G_6}
\int d^2\sigma \sqrt{g_{\sigma}} =
\frac{2 \pi r_c^2}{G_6} \,\, .
\ee
Hence, by using the field configuration given in Eq.~(\ref{FC}), we
can re-write the action in Eq.~(\ref{action}) as follows
\be
S= \frac{1}{G_4} \int d^4 x \sqrt{g_4} \Big\{ e^{2f}\big[ R_1 +
e^{-2f} R_2 + 2 (\partial_\mu f)^2 - (\partial_\mu \phi)^2 \big]  -
\frac{\xi}{G_6} e^{2f+\phi} - \frac{G_6 b^2}{r_c^4}\, e^{ -2f-\phi}
\Big\} \,.
\label{action4string}
\ee
Let us consider now a rescaling of the metric of $N_1$:
$\hat{g}_{\mu\nu}\equiv e^{2f} g_{\mu\nu}$ and
$\sqrt{\hat{g}_4}=e^{4f} \sqrt{g_4}.$ Such a transformation brings
the theory into the Einstein conformal frame where the action given in
Eq.~(\ref{action4string}) takes the form
\be S= \frac{1}{G_4} \int d^4
x \sqrt{\hat{g}_4} \Big[ R[\hat{g}_4] -4 (\partial_\mu f)^2 -
(\partial_\mu \phi)^2 - \frac{\xi}{G_6} e^{-2f+\phi}-\frac{G_6 b^2}{r_c^4}\,
e^{-6f-\phi} + e^{-4f}R_2 \Big].
\label{action4}
\ee
The four dimensional Lagrangian is then
\beq
L= \frac{\sqrt{g}}{G_4}\,\Big[ R- 4 (\partial_\mu f)^2 - (\partial_\mu
\phi)^2 - V(f,\phi)  \Big],
\label{finallagran}
\eeq
with
\beq
V(f,\phi)\equiv \frac{\xi}{G_6}e^{-2f+\phi} + \frac{G_6 b^2}{r_c^4}\,
e^{-6f-\phi} - e^{-4f}R_2 \,\,,
\eeq
where to simplify the notation we have defined: $g\equiv\hat{g}_4$
and $R\equiv R[\hat{g}_4]$.

Let us now define a new orthogonal basis, $X \equiv (\phi+2f)/\sqrt G_4$
and $Y \equiv (\phi-2f)/\sqrt G_4$, so that the kinetic energy terms in
the Lagrangian are both canonical, i.e.,
\beq
L= \sqrt{g}\left[\frac{R}{G_4} -\frac{1}{2}(\partial X)^2
-\frac{1}{2}(\partial Y)^2 - \tilde V(X,Y)  \right],
\label{ll}
\eeq
where the potential $\tilde V(X,Y) \equiv V( f, \phi)/G_4$ can be re-written (after some elementary algebra) as~\cite{Vinet:2005dg}
\beq
\tilde V(X,Y)= \frac{e^{\sqrt{G_4} Y}}{G_4} \left[ \frac{G_6
b^2}{r_c^4}e^{-2\sqrt{G_4} X} - R_2 e^{-\sqrt{G_4} X} +\frac{\xi}{G_6}
\right] \,\, .
\label{potential}
\eeq

The field equations are
\bea
 R_{\mu\nu} - \frac{1}{2}g_{\mu\nu}R & = &
    \frac{G_4}{2} \left[\left(\partial_\mu X \partial_\nu X -\frac{g_{\mu\nu}}{2}\, \partial_\eta X \, \partial^\eta X \right) \right. \nonumber \\
 & + &
\left. \left(
\partial_\mu Y \partial_\nu Y -\frac{g_{\mu\nu}}{2} \,\partial_\eta Y \, \partial^\eta Y \right) -g_{\mu\nu} \tilde V(X,Y)\right] \, ,
\label{einsteinb}
\eea
$\Box  X = \partial_X \tilde V,$ and $\Box  Y = \partial_Y \tilde V.$
In order to allow for a dS era, we assume that the metric
takes the form
\beq
ds^2= -dt^2 + e^{2h(t)}d\vec{x}^{\,2}  ,
\label{refc}
\eeq
and that $X$ and $Y$ depend only on the time coordinate, i.e., $X=X(t)$
and $Y=Y(t)$.
Then, the equations of motion for $X$ and $Y$ can be written as
\bea
\ddot X + 3 \dot h\dot X=  - \partial_X \tilde V
\eea
and
\begin{equation}
 \ddot Y + 3 \dot h\dot Y= - \partial_Y \tilde V,
\label{eqy}
\end{equation}
whereas the  only two independent components
of Eq.~(\ref{einsteinb}) are
\begin{equation}
\dot h^2 = \frac{G_4}{6}\left[\frac{1}{2}(\dot X^2 + \dot Y^2) + \tilde V(X,Y)\right]
\label{hubble}
\end{equation}
and
\begin{equation}
 2 \ddot h+3\dot h^2 =  \frac{G_4}{2}\left[-\frac{1}{2}( \dot X^2 +  \dot Y^2) + \tilde V(X,Y)\right]\, .
\end{equation}

The terms in the square brackets in Eq.~(\ref{potential}) take the
form of a quadratic function of $e^{-\sqrt{G_4}\,X}.$ This function
has a global minimum at $ e^{-\sqrt{G_4}\, X_0} = R_2\, r_c^4/(2\,
G_6\,b^2).$ Indeed, the necessary and sufficient condition for a
minimum is that $R_2 >0$, so hereafter we only consider the spherical
compactification, where $e^{-\sqrt{G_4} \, X_0} = M_{\rm Pl}^2 /(4 \pi
b^2).$ The condition for the potential to show a dS rather than an AdS
or Minkowski phase is $\xi b^2 > 1$. Now, we expand
Eq.~(\ref{potential}) around the minimum,
\begin{equation}
\tilde V(X,Y) =  \frac{e^{\sqrt{G_4}\, Y}}{G_4} \,  \left[
    {\cal K} + \frac{\overline{M_X}^2}{2}  (X-X_0)^2 +
    {\cal O} \Big((X- X_0)^3 \Big)\right] \,,
\label{minBUF}
\end{equation}
where
\begin{equation}
\overline{M_X} \equiv \frac{1}{\sqrt{\pi}\,\,\, b r_c}
\label{mxbar}
\end{equation}
and
\begin{equation}
{\cal K} \equiv \frac{M_{\rm Pl}^2}{4 \pi r_c^2 b^2} (b^2 \xi - 1) \,\, .
\label{calk}
\end{equation}
As shown by Salam-Sezgin~\cite{Salam:1984cj}, the requirements for
preserving a fraction of supersymmetry (SUSY) in spherical
compactifications to four dimension imply $b^2 \xi = 1$, corresponding
to winding number $n= \pm 1$ for the monopole configuration.
Consequently, a ($Y$-dependent) dS background can be obtained only
through SUSY breaking. For now, we will leave open the symmetry
breaking mechanism and come back to this point after our
phenomenological discussion. The $Y$-dependent physical mass of the
$X$-particles at any time is
\begin{equation}
M_X (Y)= \frac{e^{\sqrt{G_4}\, Y/2}}{\sqrt{G_4}}\ \overline{M_X}\,,
\label{mphys}
\end{equation}
which makes this a varying mass particle (VAMP)
model~\cite{Comelli:2003cv,Franca:2003zg}, although, in this case, the dependence on
the quintessence field is fixed by the theory.
The dS (vacuum) potential energy density is
\begin{equation}
V_Y = \frac{e^{\sqrt{G_4}\, Y}}{G_4}\ {\cal K} \, .
\label{rowvac}
\end{equation}
In general,
classical oscillations for the $X$ particle will occur for
\begin{equation}
M_X > H = \sqrt{\frac{G_4\rho_{\rm tot}}{3}} \,,
\label{osc}
\end{equation}
where $\rho_{\rm tot}$ is the total energy density. (This condition is well
known from axion cosmology~\cite{Preskill:1982cy}).  A necessary
condition for this to hold  can be obtained by saturating
$\rho$ with $V_{Y}$ from
Eq.~(\ref{rowvac}) and making use of Eqs.~(\ref{mxbar}) to
(\ref{osc}), which leads to $\xi b^2 <7$. Of course, as we stray from
the present into an era where the dS energy is not dominant, we
must check at every step whether the inequality (\ref{osc}) holds. If
the inequality is violated, the $X$-particle ceases to behave like
CDM.

In what follows, some combination of the parameters of the model will
be determined by fitting present cosmological data.    To
this end we assume that SM fields are confined to $N_1$ and we
denote with $\rho_{\rm rad}$ the radiation energy, with
$\rho_X$ the matter energy associated with the $X$-particles, and with
$\rho_{\rm mat}$ the remaining matter density. With this in mind,
Eq.~(\ref{eqy}) can be re-written as \be \ddot Y + 3 \,H \,\dot Y = -
\frac{\partial V_{\rm eff}}{\partial Y} \,, \ee where $V_{\rm eff}
\equiv V_Y +\rho_X$ and $H$ is defined by the Friedmann equation \be
H^2 \equiv \dot{h}^2 = \frac{1}{3 M_{\rm Pl}^2} \, \left[
  \frac{1}{2}\, \dot Y^2 + V_{\rm eff} + \rho_{\rm rad} + \rho_{\rm
    mat} \right] \,\,.  \ee (Note that the matter energy associated to
the $X$ particles is contained in $V_{\rm eff}$.)

It is more convenient to consider the evolution in $u \equiv - \ln (1 +z),$
where $z$ is the redshift parameter.
As long as the oscillation condition is fulfilled, the VAMP CDM energy
density is given in terms of the $X$-particle number density
$n_X$~\cite{Hoffman:2003ru}
\begin{equation}
\rho_X(Y,u) = M_X(Y)\ n_X(u)\\[.1in]
= C \ e^{\sqrt{G_4} Y/2}\ e^{-3u} \,\,,
\label{rowx}
\end{equation}
where $C$ is a constant to be determined by fitting to data.
Along with Eq.~(\ref{rowvac}), these define for us the effective
($u$-dependent) VAMP potential
\be V_{\rm eff}(Y,u)\equiv V_Y +\rho_X=  A\ e^{\sqrt{G_4}Y} +
C \
e^{\sqrt{G_4}Y/2}\ e^{-3u}\, ,
\label{veff}
\ee
where a $A$ is just a constant given in terms of model parameters through
Eqs.~(\ref{minBUF}) and (\ref{calk}).

Hereafter we adopt natural units, $M_{\rm Pl} = 1.$ Denoting by a
prime derivatives with respect to $u,$ the equation of motion for $Y$
becomes \be \frac{Y^{\prime\prime}}{1 - \,
  Y^{\prime^2}\!/6} \,\, + 3 \,Y^\prime +
\frac{\partial_u\rho\,\, \,Y^\prime/2 \,\,+ \,3 \,\,\partial_Y V_{\rm
    eff}}{\rho} = 0\,,
\label{motion}
\ee
where $\rho = V_{\rm eff} + \rho_{\rm rad} + \rho_{\rm mat}.$
Quantities of importance are the dark energy density
\be \rho_{Y} =
\frac{1}{2}\, H^2 \,Y^{\prime 2} + V_Y \,,
\ee
generally expressed in
units of the critical density ($\Omega\equiv\rho/\rho_{\rm c}$)
\be
\Omega_Y = \frac{\rho_Y}{3 H^2}\,,
\ee
and the Hubble parameter
\be
H^2 = \frac{\rho}{3 - Y^{\prime 2}/2}\,\,.
\ee
The equation of state
is
\be w_Y = \left[\frac{H^2 \,\,Y^{\prime 2}}{2} - V_Y\right]\,
\left[\frac{H^2 \,\,Y^{\prime 2}}{2}+ V_Y\right]^{-1}\,.
\ee
We pause to note that the exponential potential $V_Y
\sim e^{\lambda Y/M_{\rm Pl}},$ with $\lambda =\sqrt{2}.$
Asymptotically,
this represents the crossover situation with 
$w_Y = -1/3$~\cite{Copeland:1997et},
implying expansion at constant velocity. Nevertheless, we will find that there
is a brief period encompassing the recent past $(z\leq 6)$ where there has
been significant acceleration.

Returning now to the quantitative analysis, we take $\rho_{\rm mat} =
B e^{-3u}$ and $\rho_{\rm rad} = 10^{-4} \ \rho_{\rm mat}$ $ e^{-u}\
f(u)$ where $B$ is a constant and $f(u)$ parameterizes the
$u$-dependent number of radiation degrees of freedom.\footnote{This
  assumption will be justified {\em a posteriori} when we find that
  $\rho_X \ll \rho_{\rm mat}.$} In order to interpolate the various
thresholds appearing prior to recombination (among others, QCD and
electroweak), we adopt a convenient phenomenological form
$f(u)=\exp(-u/15)$~\cite{Anchordoqui:2003ij}.  We note at this point
that solutions of Eq.~(\ref{motion}) are independent by an overall
normalization for the energy density. This is also true for the
dimensionless quantities of interest $\Omega_Y$ and $w_Y.$

\begin{figure}[!thb]
\postscript{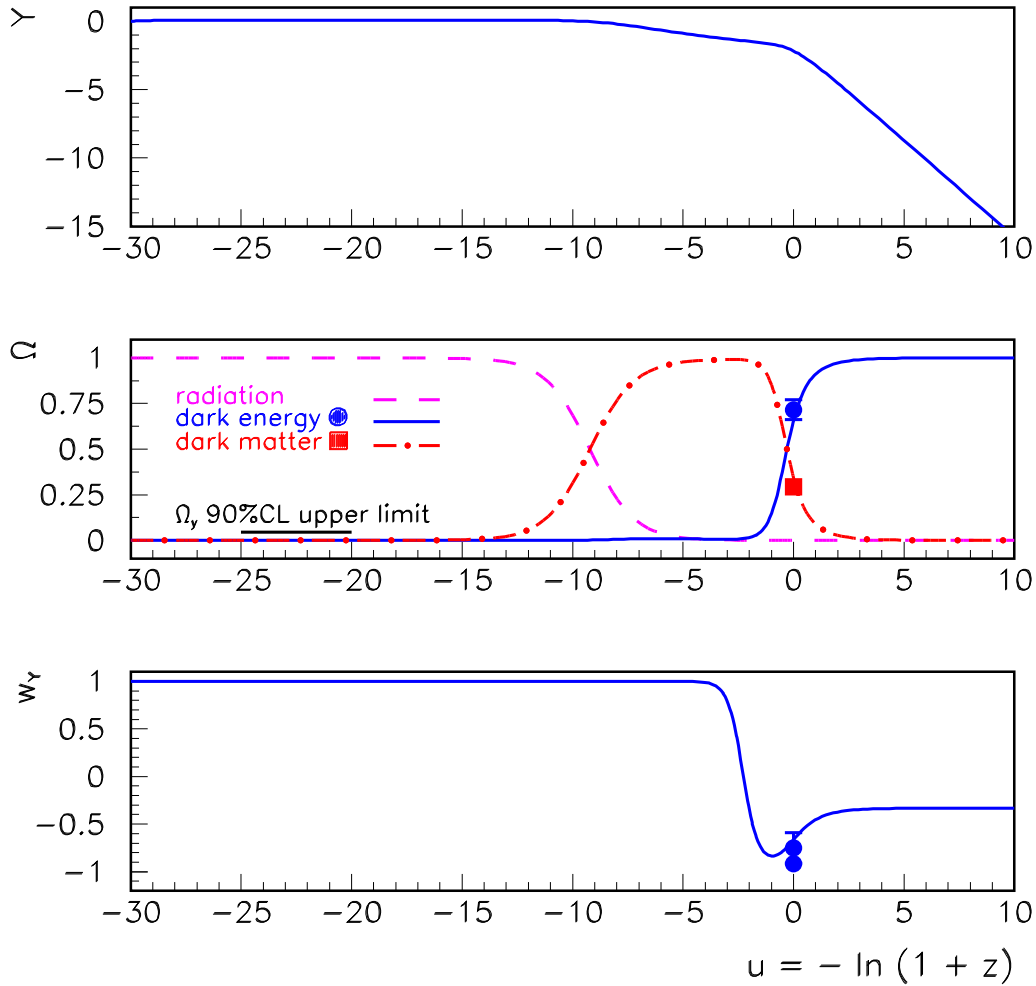}{0.9}
\caption[Eyeball fit of cosmological data using Salam-Sezgin supergravity] {The upper panel shows the evolution of $Y$ as a function of
  $u$. Today corresponds to $z=0$ and for primordial nucleosynthesis
  $z \approx 10^{10}.$ We set the initial conditions $Y (-30) = 0$ and
  $Y'(-30) = 0.08;$ we take $A : B : C = 11 : 0.3 : 0.1.$
  The second panel shows the evolution of
  $\Omega_Y$ (solid line), $\Omega_{\rm mat}$ (dot-dashed line), and
  $\Omega_{\rm rad}$ (dashed line) superposed over experimental best
  fits from SDSS and WMAP
  observations~\cite{Tegmark:2003ud,Spergel:2006hy}. The curves are
  not actual fits to the experimental data but are based on the
  particular choice of the $Y$ evolution shown in the upper panel,
  which provides eyeball agreement with existing astrophysical
  observations.  The lower panel shows the evolution of the equation
  of state $w_Y$ superposed over the best fits to WMAP + SDSS data
  sets and WMAP + SNGold~\cite{Spergel:2006hy} . The solution of the
  field equations is consistent with the requirement from primordial
  nucleosynthesis, 
$\Omega_Y < 0.045$ (90\%CL)~\cite{Olive:1999ij,Bean:2002sm}; it
  also shows the established radiation and matter dominated epochs,
  and at the end shows an accelerated dS era~\cite{Anchordoqui:2007sb}.}
\label{sancle}
\end{figure}

With these forms for the energy densities, Eq.~(\ref{motion}) can be
integrated for various choices of $A,$ $B,$ and $C$, and initial
conditions at $u=-30.$ We take as initial condition $Y(-30) = 0$.
Because of the slow variation of $Y$ over the range of $u,$ changes in
$Y(-30)$ are equivalent to altering the quantities $A$ and
$C$~\cite{LopesFranca:2002ek}. In accordance to equipartition
arguments~\cite{LopesFranca:2002ek,Steinhardt:1999nw} we take $Y'(-30)
= 0.08.$ Because the $Y$ evolution equation depends only on energy
density ratios, and hence only on the ratios $A:B:C$ of the previously
introduced constants, we may, for the purposes of integration and
without loss of generality, arbitrarily fix $B$ and then scan the $A$
and $C$ parameter space for applicable solutions. In Fig.~\ref{sancle}
we show a sample qualitative fit to the data. It has the property of
allowing the maximum value of $X$-CDM (about 7\% of the total dark
matter component) before the fits deviate unacceptably from data.

It is worth pausing at this juncture to examine the consequences of
this model for variation in the fine structure constant and long range
forces. Specifically, excitations of the electromagnetic field on
$N_1$ will, through the presence of the dilaton factor in
Eq.~(\ref{action}), seemingly induce variation in the electromagnetic
fine structure constant $\alpha_{\rm em} = e^2/4\pi$, as well as a
violation of the equivalence principle through a long range coupling
of the dilaton to the electromagnetic component of the stress tensor.
We now show that these effects are extremely negligible in the present model.
First, it is easily seen using Eqs.~(\ref{action}) and (\ref{metric})
together with Eqs.~(\ref{007})-(\ref{potential}), that the
electromagnetic piece of the lagrangian as viewed from $N_1$ is 
\begin{equation}
 {\cal L}_{\rm em} = -\frac{2\pi}{4} e^{- {\sqrt{G_4}X}} 
\widetilde{f}_{\mu\nu}^2 \,\,,
\end{equation}
where $\widetilde{f}_{\mu \nu}$ denotes a quantum fluctuation of the
electromagnetic $U(1)$ field. (Fluctuations of the $U(1)$ background
field are studied in the next section.) At the equilibrium value
$X=X_0$, the exponential factor is
\begin{equation}
 e^{- {\sqrt{G_4}X_0}} =
 \frac{M_{\rm Pl}^2}{4\pi b^2} \,,
\end{equation}
so that we can identify the electromagnetic coupling $(1/e^2) \simeq
M_{\rm Pl}^2/ b^2$. This shows that $b\sim M_{\rm Pl}$.  We can then
expand about the equilibrium point, and obtain an additional factor of
$(X-X_0)/M_{\rm Pl}$. This will do two things~\cite{Carroll:1998zi}:
$(a)$ At the classical level, it will induce a variation of the
electromagnetic coupling as $X$ varies, with $\Delta\alpha_{\rm
  em}/\alpha_{\rm em} \simeq (X-X_0)/M_{\rm Pl}$; and $(b)$~at the quantum
level, exchange of $X$ quanta will induce a new force through coupling
to the electromagnetic component of matter.

Item $(b)$ is dangerous if the mass of the exchanged quanta are small,
so that the force is long range.  This is not the case in the present
model: from Eq.~(\ref{minBUF}) the $X$ quanta have mass of ${\cal
  O}({\overline M}_X M_{\rm Pl}) \sim M_{\rm Pl}/(r_c b)$, so that if
$r_c$ is much less than ${\cal O}({\rm cm}),$ the forces will play
no role in the laboratory or cosmologically.

As far as the variation of $\alpha_{\rm em}$ is concerned, we find that
$\rho_X/\rho_{\rm mat} = (C/B) e^{Y/\sqrt{2}},$ so that
\begin{eqnarray}
\rho_X &\simeq& 3\times 10^{-120} e^{-3u}M_{\rm Pl}^4 e^{Y/\sqrt{2}}
\nonumber \\
&=& \frac{1}{4} \overline M _X^2 (X-X_0)^2 e^{Y \sqrt{2}}M_{\rm Pl}^2 \,\,.
\end{eqnarray}
This then gives,
\begin{equation}
 \sqrt{\langle(X-X_0)^2 \rangle }\equiv \Delta X_{\rm rms}  \approx 10^{-60}
e^{-3u/2} M_{\rm Pl} e^{Y/(2\sqrt{2})}/\overline M _X \,\, .
\end{equation}
During the radiation era, $Y\simeq \ {\rm const} \ \simeq 0$ (see
Fig.~\ref{sancle}), so that during nucleosynthesis ($u\simeq -23)$
$\Delta X_{\rm rms}/M_{\rm Pl} \simeq 10^{-45}/\overline M _X,$
certainly no threat. It is interesting that such a small value can be
understood as a result of inflation: from the equation of motion for
the $X$ field, it is simple to see that during a dS era with
Hubble constant $H$, the amplitude $\Delta X_{\rm rms}$ is damped as
$e^{-3Ht/2}$. For 50~$e$-foldings, this represents a damping of
$10^{32}.$ In order to make the numbers match (assuming a
pre-inflation value $\Delta X_{\rm rms}/M_{\rm Pl}\sim 1$), an
additional damping of $\sim 10^{13}$ is required from reheat
temperature to primordial
nucleosynthesis. With the $e^{-3u/2}$ behavior, this implies a low
reheat temperature, about $10^6$ GeV.
Otherwise, one may just assume an additional fine-tuning of the
initial condition on $X$.

As mentioned previously, the solutions of Eq.~(\ref{motion}), as well
as the quantities we are fitting to ($\Omega_Y$ and $w_Y$), depend
only on the ratios of the energy densities. From the eyeball fit in
Fig.~\ref{sancle} we have, up to a common constant, $\rho_{\rm
  ordinary\ matter} \equiv \rho_{\rm mat} \propto 0.3\ e^{-3u}$ and $
V_Y \propto 11\ e^{\sqrt{2}Y}.$ We can deduce from these relations
that \be \frac{V_Y({\rm now})}{\rho_{\rm mat}({\rm now})} =
\frac{11}{0.3}\ e^{\sqrt{2}Y({\rm now})} \simeq 36\ e^{\sqrt{2}Y({\rm
    now})} \,\,.
\label{ratio}
\ee
Besides, we know that $\rho_{\rm mat}({\rm now}) \simeq 0.3
\rho_c({\rm now})\simeq 10^{-120}\
M_{\rm Pl}^4.$ Now, Eqs.~(\ref{minBUF}) and (\ref{calk}) lead to
\be
 V_Y({\rm now}) = e^{\sqrt{2}Y({\rm now})}\
\frac{M_{\rm Pl}^4}{8\pi\ r_c^2\ b^2}\ (b^2\xi -1)
\label{theory}
\ee
so that from Eqs.~(\ref{ratio}) and (\ref{theory}) we obtain
\be
\frac{1}{8\pi\ r_c^2\ b^2}\ (b^2\xi -1) \simeq 10^{-119} \,\,.
\label{finetuning}
\ee It is apparent that this condition cannot be naturally
accomplished by choosing large values of $r_c$ and/or $b.$ There
remains the possibility that SUSY breaking~\cite{Aghababaie:2003wz} or
non-perturbative effects lead to an exponentially small deviation of
$b^2 \xi$ from unity, such that $b^2 \xi = 1 + {\cal O}
(10^{-119})$.\footnote{ Before proceeding, we remind the reader that
  the requirements for preserving a fraction of SUSY in spherical
  compactifications to four dimensions imply $b^2 \xi = 1$,
  corresponding to the winding number $n = \pm 1$ for the monopole
  configuration. In terms of the Bohm-Aharonov argument on phases,
  this is consistent with the usual requirement of quantization of the
  monopole. The SUSY breaking has associated a non-quantized flux of
  the field supporting the two sphere. In other words, if we perform
  a  Bohm-Aharonov-like interference experiment, some phase change
  will be detected by a $U(1)$ charged  particle that circulates
  around the associated Dirac string. The quantization of fluxes
  implied the unobservability of such a phase, and so in our
  cosmological set-up, the parallel transport of a fermion will be
  slightly  path dependent. One possibility is that the non-compact
  $\rho$ coordinate (in the uplift to ten dimensions, see Appendix B) is
  the direction in which the Dirac string exists. Then the cutoff
  necessary on the physics at large  $\rho$  will introduce a slight
  (time-dependent) perturbation on the flux quantization condition.}
Since a deviation of $b^2\xi$ from unity involves a breaking of
supersymmetry, a small value for this dimensionless parameter, perhaps
$(1\ {\rm TeV}/M_{\rm Pl})^2\sim 10^{-31}$, can be expected on the
basis of 't Hooft naturalness.  It is the extent of the smallness, of
course, which remains to be explained.

\section{Fluctuations in the Background Configuration}

In this section we study the quantum fluctuations of the $U(1)$ field
associted to the background configuration.  We start by considering
fluctuations of the background field $A_M^0$ in the 4 dimensional
space, i.e,
\begin{equation}
A_M \to A_M^0 + \epsilon \ a_M \,,
\label{amu}
\end{equation}
where $A_M^0 = 0$ if $M \neq \varphi$ and $a_M = 0$ if 
$M = \vartheta, \varphi.$ The fluctuations on $A_M^0$ lead to
\begin{equation}
F_{MN} \to F_{MN}^0 + \epsilon \ f_{M N} \,\, .
\label{fmunu}
\end{equation}
Then,
\begin{equation}
F_{MN} F^{MN} =  g^{ML} \ g^{NP} [F_{MN}^0 F^0_{LP} + \epsilon \ F_{MN}^0 \ f_{LP}  + 
\epsilon^2 f_{MN}\ f_{LP}] \  .
\label{Fdos}
\end{equation}
The second term vanishes and the first and third terms are nonzero
because $F_{MN}^0 \neq 0$ in the compact space and $f_{MN} \neq 0$ in
the 4 dimensional space.  If the Kalb-Ramond potential $B_{NM} = 0$,
then the 3-form field strength can be written as
\begin{equation}
G_{MNP} =\kappa A_{[M} \ F_{NP]}=
\frac{\kappa}{3!}\ [A_M \ F_{NP} + A_P \ F_{MN} - A_N \ F_{MP}]\, .
\end{equation}
Now we introduce notation of differential forms, in which
the usual Maxwell field and field strength read
\beq
A_1= A_M dx^M\;\; {\rm and} \;\; F_2= F_{MN} \, dx^M \wedge dx^N\,\,;
\eeq
respectively. (Note that $dx^M \wedge dx^N$ is antisymmetrized by 
definition.) With this in mind, the 3-form reads
\beq
G_3= \kappa A_1 \wedge F_2= \kappa A_M F_{NP} \  dx^M \wedge dx^N
\wedge dx^P \,\, .
\label{nes}
\eeq
Substituting Eqs.~(\ref{amu}) and (\ref{fmunu}) into Eq.~(\ref{nes}) we obtain
\beq
G_3=\kappa \Big[ (A_M^0 + \epsilon a_M)(F_{NP}^0 + \epsilon f_{N P}) \
dx^M \wedge dx^N \wedge dx^P\Big] \,\, .
\label{1}
\eeq
The background fields read
\beq
A_1^0= b\, \cos\vartheta \, d\varphi,\;\;\; F_2^0= -b\, 
\sin\vartheta\, d\vartheta \wedge d\varphi \,\,,
\label{ocho}
\eeq 
and the fluctuations on the probe brane become
\beq
a_1= a_\mu dx^\mu,\;\;\; f_2= f dx^\mu \wedge dx^\nu,\;\;{\rm with}\;\; f= 
\partial_\mu a_\nu -\partial_\mu a_\nu \, .
\eeq
All in all,
\begin{eqnarray}
\frac{G_3}{\kappa} & = & A^{0}_\varphi F^{0}_{\vartheta\varphi} \ 
d\varphi\wedge 
d\vartheta \wedge d\varphi +\epsilon A_\varphi^0 f_{\mu \nu} \ d\varphi 
\wedge dx^\mu 
\wedge dx^\nu + \epsilon F_{\vartheta\varphi}^0 a_\mu \ d\vartheta\wedge 
d\varphi\wedge 
dx^\mu \nonumber \\
 & + & \epsilon^2 a_\mu f_{\zeta \nu} dx^\mu\wedge dx^\zeta \wedge dx^\nu \ .
\label{2} 
\end{eqnarray}
Using Eq.~(\ref{ocho}) and the antisymmetry of the wedge product,
Eq.~(\ref{2}) can be re-written as \beq
\frac{G_3}{\kappa}=\epsilon\Big[b\cos\vartheta f_{\mu \nu}d\varphi
\wedge dx^\mu \wedge dx^\nu - b a_\mu \sin\vartheta d\vartheta\wedge
d\varphi\wedge dx^\mu +\epsilon a_\mu f_{\zeta \nu} dx^\mu \wedge
dx^\zeta \wedge dx^\nu\Big] \, .
\label{3}
\eeq
From the metric
\beq
ds^2= e^{2\alpha}dx_{4}^2 + e^{2\beta} 
(d\vartheta^2 +\sin\vartheta^2 d\varphi^2)
\eeq
we can write the vielbeins 
\bea
& & e^{a}= e^\alpha dx^a,\;\;\; e^\vartheta= e^\beta d\vartheta,\;\;\; e^\varphi= 
e^\beta\sin\vartheta d\varphi,\nonumber\\
& & dx^a= e^{-\alpha} e^a,\;\;\; d\vartheta = e^{-\beta}e^\vartheta,\;\;\; 
d\varphi=\frac{e^{-\beta}}{\sin\vartheta} e^\varphi
\label{4}
\eea where $\beta \equiv f + \ln r_c.$ (Lower latin indeces from the
beginning of the alphabet indicate coordinates associted to the four
dimensional Minkowski spacetime with metric $\eta_{ab}$.)
Substituting into Eq.~(\ref{3}) we obtain 
\beq
\frac{G_3}{\kappa}=\epsilon\Big[b\frac{\cos\vartheta}{\sin\vartheta}
 e^{-2\alpha -\beta}  f_{ab}  e^\varphi \wedge e^a \wedge e^b - b
e^{-\alpha-2\beta} a_a e^\vartheta\wedge e^\varphi\wedge
e^{a}+\epsilon e^{-3\alpha}a_a f_{cb} e^{a}\wedge e^{c} \wedge e^{b}
\Big] \, ,
\label{5}
\eeq
where $f_{ab} = \partial_a a_b - \partial_b a_a$.
Because the three terms are orthogonal to each 
other, a straightforward calculation leads to
\beq
G^2_3= \kappa^2 \epsilon^2 (b^2\,\cot^2\vartheta \, e^{-4\alpha -2\beta} f_{ab}^2 + b^2 
e^{-2\alpha-4\beta} a_a^2)+ {\cal O}(\epsilon^4) \,.
\eeq
Then, the 5th term in Eq.~(\ref{action}) can be written as
\begin{eqnarray}
S_{G_3}  & = &- \frac{1}{2G_6}\, \int d^4 x \frac{G_6}{6}e^{4\alpha+2\beta}\sqrt{\eta_4}
e^{-2\phi} 
\int d\vartheta d\varphi \sin\vartheta \left[ \Big( \kappa^2 \epsilon^2 b^2\,\,\cot^2\vartheta  
e^{-4\alpha-2\beta} \Big) f_{ab}^2 \right. \nonumber \\
& + & \left. \Big( \kappa^2 \epsilon^2  b^2
e^{-2\alpha-4\beta}  \Big) a_a^2  \right] \, ,
\label{7}
\end{eqnarray}
whereas the contribution from the 4th term
in Eq.~(\ref{action}) can be computed from
Eq.~(\ref{Fdos}) yielding 
\bea
S_{F_2} &=&- \frac{1}{2G_6}\, \int d^4x \sqrt{\eta_4} 2 \pi e^{2\beta-\phi} G_6 \epsilon^2 
f_{ab}^2 \nonumber \\
&=&-\int d^4x \sqrt{\eta_4}  \pi e^{2f-\phi}r_c^2 \epsilon^2 f_{ab}^2 \,\, .
\eea
Thus,
\begin{equation}
S_{G_3} + S_{F_2} = - \int d^4x \left[\frac{1}{4\, g^2} f_{ab}^2 + \frac{m^2}{2}\, a_a^2 \right] \,,
\label{choco}
\end{equation}
where the four dimensional effective coupling  and the effective mass 
are of the form
\begin{equation}
\frac{1}{g^2}= 4\,  \epsilon^2 \sqrt{\eta_4} \left[
 \pi e^{2f - \phi} r_c^2 +
\frac{1}{12} \kappa^2  b^2 e^{-2\phi} \int d\vartheta d\varphi 
\sin\vartheta \cot^2\vartheta \right] \to \infty 
\end{equation}
and
\begin{equation}
m^2 = \frac{2}{3}\pi \k^2 b^2 \epsilon^2 e^{2\alpha-2\beta-2\phi} \, .
\label{9}
\end{equation}
For the moment we let $\int d\vartheta d\varphi \sin\vartheta
\cot^2\vartheta = N$, where eventually we set $N \to \infty.$ Now to
make quantum particle identification and coupling, we carry out the
transformation $a_a \to g \hat a_a$.\footnote{This is because the
  definition of the propagator with proper residue for correct Feyman
  rules in perturbation theory, and therefore also the couplings,
  needs to be consistent with the form of the Hamiltonian $= \sum_k
  \omega(k) a_k^\dagger a_k,$ with $[a,\, a^\dagger] =1$. This in turn
  implies that the kinetic term in the Lagrangian has the canonical
  form, $(1/4) \hat f_{ab}^2,$ with the usual expansion of the vector
  field $a_a.$} This implies that the second term in the right hand
side of Eq.~(\ref{choco}) vanishes, yielding
\begin{equation}
 f_{ab}  = \partial_a (g \hat a_b) - \partial_b (g \hat a_a) 
  =  \partial_a g \, \hat a_b - \partial_b g\, \hat a_a + g \, 
  \partial_a   \hat a_b - g\, \partial_b \hat a_a 
  =  g \hat f_{ab} + \hat a \wedge dg
\end{equation}
and consequently to leading order in $N$
\begin{equation}
\frac{1}{g^2} \, f_{ab}^2 = \frac{1}{g^2} [g^2 \hat f_{ab}^2 + (\hat a 
\wedge dg)^2 + 2\,g \, \hat a_b \,\, \hat f^{ab}\, \partial_a g] \,\, .
\end{equation}
If the coupling depends only  on the time variable, 
\begin{equation}
\frac{1}{g^2} \, f_{ab}^2  \to  \hat f_{ab}^2 + 
\left(\frac{\dot g}{g}\right)^2 \, \hat a_a^2  + 2 \, \frac{\dot g}{g} \,\, 
\hat a_i \,\, \hat f^{ti} \,\,,
\end{equation}
where $\dot g = \partial_t g$ and lower latin indices from the middle
of the alphabet refer to the brane space-like dimensions.  If we
choose a time-like gauge in which $a_t = 0,$ then the term $(\dot
g/g)\, \hat a_i \, \hat f^{ti}$ can be written as $(1/2) (\dot g/g)
(d/dt) (\hat a_i)^2,$ which after an integration by parts gives $-
(1/2) [(d/dt) (\dot g/g)] \hat a_i^2$; with $g \sim e^{-\phi},$ the factor
in square brackets becomes $- \ddot \phi.$ Since $\phi =\sqrt{G_4} (X + Y),$
the rapidly varying $\ddot X$ will average to zero, and one is left
just with the very small $\ddot Y$, which is of order Hubble square. 
For the term $(\dot g/g)^2 (a_i)^2,$ the term $(\dot X)^2$ also averages to 
order Hubble square, implying that the induced mass term is of horizon size. 
These ``paraphotons''
carry new relativistic degrees of freedom, which could in turn modify
the Hubble expansion rate during Big Bang nucleosynthesis (BBN). Note,
however, that these extremely light gauge bosons are thought to be
created through inflaton decay and their interactions are only
relevant at Planck-type energies. Since the quantum gravity era, all
the paraphotons have been redshifting down without being subject to
reheating, and consequently at BBN they only count for a fraction of an
extra neutrino species in agreement with observations.

%% file: chap_jets.tex
The CERN's LHC is the greatest basic science
endeavor in history. Spectacular physics results are expected to
follow in short order once it turns on this year. The LHC will push
nucleon-nucleon center-of-mass energies up to $\sqrt{s} = 5.5~{\rm
  TeV}$ for Pb-Pb collisions, and $\sqrt{s} = 14~{\rm TeV}$ for $pp$
collisions. The ALICE detector will observe the very messy debris of
heavy ion collisions, whereas the ATLAS and CMS detectors will observe
the highest-energy particle collisions produced by the accelerator.
The LHC will probe deeply into the sub-fermi distances, committing to
careful searches for new particles and interactions at the TeV scale.

At the time of its formulation and for years thereafter, superstring
theory was regarded as a unifying framework for Planck-scale quantum
gravity and TeV-scale SM physics. Important advances were fueled by
the realization of the vital role played by D-branes~\cite{joe} in
connecting string theory to phenomenology~\cite{reviews}. This has
permitted the formulation of string theories with compositeness
setting in at TeV scales~\cite{Antoniadis:1998ig,Lykken:1996fj} and
large extra dimensions. There are two paramount phenomenological
consequences for TeV scale D-brane string physics: the emergence of
Regge recurrences at parton collision energies $\sqrt{\hat s} \sim
{\rm string\ scale} \equiv M_s;$ and the presence of one or more
additional $U(1)$ gauge symmetries, beyond the $U(1)_Y$ of the SM.
The latter follows from the property that the gauge group for open
strings terminating on a stack of $N$ identical D-branes is $U(N)$
rather than $SU(N)$ for $N>2.$ (For $N=2$ the gauge group can be
$Sp(1)$ rather than $U(2)$.) In this chapter we exploit both these
properties in order to obtain a ``new physics'' signal at the LHC
which, if traced to low scale string theory, could with 100~fb$^{-1}$
of data probe deviations from the SM physics at a $5\sigma$
significance for $M_s$ as large as 2.3~TeV.

\section{Perturbative D-brane Models}

The concept of D-branes was introduced in the late
80's~\cite{Dai:1989ua}.  They are described as a geometric locus where
strings can end. In a quantum gravity theory, like string theory, any
defect or extended object in spacetime can bend and it will
consequently have excitations.  For D-branes, the excitations are the
open string attached to them. Furthermore, one can have various
D-branes on top of one another. In these situations one needs to
consider open strings with Chan-Paton indices~\cite{Paton:1969je} on
them, and thus one has continuous gauge symmetries associated to the
ends of the string.  The allowed gauge groups in these D-brane
constructions are those that can have a large $N$ limit: $U(N),\ SO(N),\
Sp(N)$.  Besides, each end of the string ends carries a fundamental
charge with respect to the stack of branes on which it ends. Hence,
any open string will carry the quantum numbers associated to some type
of bifundamental representation. These novel constructions provide a
framework for particle physics on a brane, yielding a possible
realization of the SM within string theory.

\begin{figure}[tbp]
\postscript{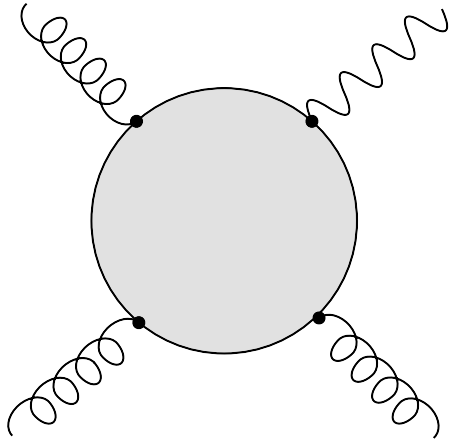}{0.4}
\caption[Open string scattering diagram]{Open string scattering diagram for $gg \to g \gamma$.  The
  dots represent vertex insertions of gauge bosons on the boundary of
  the world sheet.}
\label{diag}
\end{figure}

To describe the field theory degrees of freedom it is convenient to
introduce a graphic notation, generally referred to as quivers or moose
diagrams. The gauge degrees of freedom (brane stacks) are then
described by nodes in a graph. The open string particles are given by
edges connecting two vertices and arrows that dictate if the
corresponding end of the string is fundamental or anti-fundamental.
One should also label the edges according to the other quantum numbers
that the particles carry. In the perturbative regime, where the low
energy dynamics is given in terms of open strings alone (all other
non-perturbative states are heavy), the interactions are generated by
disc diagrams (a relevant example is pictured in Fig.~\ref{diag}).
These are single traces of fields. If a vertex has $n+2$ particles
attached to it, it will appear with a coupling constant dependence of
$g^{n}$, where $g$ is the open string coupling constant.

To develop our program in the simplest way, we will work within the
construct of a minimal model in which we consider scattering processes
which take place on the (color) $U(3)$ stack of D-branes. In the
bosonic sector, the open strings terminating on this stack contain, in
addition to the $SU(3)$ octet of gluons, an extra $U(1)$ boson
($C_\mu$, in the notation of~\cite{Berenstein:2006pk}), most simply
the manifestation of a gauged baryon number symmetry. The $U(1)_Y$
boson $Y_\mu$, which gauges the usual electroweak hypercharge
symmetry, is a linear combination of $C_\mu$, the $U(1)$ boson $B_\mu$
terminating on a separate $U(1)$ brane, and perhaps a third additional
$U(1)$ (say $W_\mu$) sharing a $U(2)$ brane to which are also a
terminus for the $SU(2)_L$ electroweak gauge bosons $W_\mu^a.$ Thus,
critically for our purposes, the photon $A_\mu$, which is a linear
combination of $Y_\mu$ and $W^3_\mu$, {\em will participate with the
  gluon octet in (string) tree level scattering processes on the color
  brane, processes which in the SM occur only at one-loop level.} Such
a mixing between hypercharge and baryon number is a generic property
of D-brane quivers, see {\it e.g}.\
Refs.~\cite{Berenstein:2006pk,ant,bo}. The vector boson $Z'_\mu$,
orthogonal to the hypercharge, must grow a mass $M_{Z'}$ in order to
avoid long range forces between baryons other than gravity and Coulomb
forces. The anomalous mass growth allows the survival of global baryon
number conservation, preventing fast proton
decay~\cite{Ghilencea:2002da}.

The processes we consider (at the parton level) are $gg\rightarrow
g\gamma$ and $gg \rightarrow \gamma \gamma$, where $g$ is an $SU(3)$
gluon and $\gamma$ is the photon. As explicitly calculated below,
these will occur at string disk (tree) level, and will be manifest at
the LHC as a non-SM contribution to $pp\rightarrow \gamma +\ {\rm
  jet}$ and $pp \rightarrow \gamma \gamma$.  A very important property
of string disk amplitudes is that they are completely
model-independent; thus the results presented below are robust,
because {\em they hold for arbitrary compactifications of superstring
  theory from ten to four dimensions, including those that break
  supersymmetry}.  The SM background for these signals originates in
the parton tree level processes $g q \rightarrow \gamma q,\ g\bar
q\rightarrow \gamma\bar q,\ q\bar q\rightarrow \gamma g,\ {\rm and} \
q\bar q\rightarrow \gamma \gamma $. Of course, the SM processes will
also receive stringy corrections which should be added to the pure
bosonic contribution as part of the signal.\footnote{Some qualitative
  and quantitative considerations of these processes have been discussed
  in~\cite{Burikham:2004su,Cheung:2005ig,Domokos:1998ry}.} We postpone
their evaluation until Chapter~5. Thus, the contribution from the
bosonic process calculated here is to be regarded as a lower bound to
the stringy signal. It should also be stated that, in what follows, we
do not include effects of Kaluza-Klein recurrences due to
compactification. We assume that all such effects are in the
gravitational sector, and hence occur at higher order in string
coupling~\cite{Cullen:2000ef}.

\section{The String Amplitude}

The most direct way to compute the amplitude for the scattering of
four gauge bosons is to consider the case of polarized particles
because all non-vanishing contributions can then be generated from a
single, maximally helicity violating (MHV), amplitude -- the so-called
{\it partial\/} MHV amplitude~\cite{ptmhv}.  Assume that two vector
bosons, with the momenta $k_1$ and $k_2$, in the $U(N)$ gauge group
states corresponding to the generators $T^{a_1}$ and $T^{a_2}$ (here
in the fundamental representation), carry negative helicities while
the other two, with the momenta $k_3$ and $k_4$ and gauge group states
$T^{a_3}$ and $T^{a_4}$, respectively, carry positive helicities. (All
momenta are incoming.) Then the partial amplitude for such an MHV
configuration is given by~\cite{STii,STi}
\begin{equation}
\label{ampl}
A(1^-,2^-,3^+,4^+) ~=~ 4\, g^2\, {\rm Tr}
  \, (\, T^{a_1}T^{a_2}T^{a_3}T^{a_4}) {\langle 12\rangle^4\over
    \langle 12\rangle\langle 23\rangle\langle 34\rangle\langle
    41\rangle}V(k_1,k_2,k_3,k_4)\ ,
\end{equation}
where $g$ is the $U(N)$ coupling constant, $\langle ij\rangle$ are the
standard spinor products written in the notation of
Refs.~\cite{Mangano,Dixon}, and the Veneziano
formfactor\cite{Veneziano:1968yb,Veneziano:1974dr},
\begin{equation}
\label{formf}
V(k_1,k_2,k_3,k_4)=V(s,t,u)= {\Gamma(1-s)\ \Gamma(1-u)\over
    \Gamma(1+t)}\ ,
\end{equation}
is the function of Mandelstam variables, here
normalized in the string units:
\begin{equation}
\label{mandel}
s={2k_1k_2\over M_s^2},~ t={2
  k_1k_3\over M_s^2}, ~u={2 k_1k_4 \over M_s^2}:\qquad s+t+u=0.
\end{equation}
(For simplicity we drop carets for the parton subprocess.)
Its low-energy
  expansion reads
\begin{equation}
\label{vexp}
V(s,t,u)\approx 1-{\pi^2\over 6}s\,
    u-\zeta(3)\,s\, t\, u+\dots
\end{equation}

We first consider the amplitude involving three $SU(N)$ gluons
$g_1,~g_2,~g_3$ and one $U(1)$ gauge boson $\gamma_4$ associated to
the same $U(N)$ quiver:
\begin{equation}
\label{gens}
T^{a_1}=T^a \ ,~ \ T^{a_2}=T^b\ ,~ \
  T^{a_3}=T^c \ ,~ \ T^{a_4}=Q_cI\ ,
\end{equation}
where $I$ is the $N{\times}N$ identity matrix and $Q_c$ is the
$U(1)$ charge of the fundamental representation. The $U(N)$
generators are normalized according to
\begin{equation}
\label{norm}
{\rm Tr}(T^{a}T^{b})={1\over 2}\delta^{ab}.
\end{equation}
Then the color
factor \begin{equation}\label{colf}{\rm
    Tr}(T^{a_1}T^{a_2}T^{a_3}T^{a_4})=Q_c(d^{abc}+{i\over 4}f^{abc})\ ,
\end{equation}
where the totally symmetric symbol $d^{abc}$ is the symmetrized trace,
while $f^{abc}$ is the totally antisymmetric structure constant.

The full MHV amplitude can be obtained~\cite{STii,STi} by summing
the partial amplitudes (\ref{ampl}) with the indices permuted in the
following way: \begin{equation}
\label{afull} {\cal M}(g^-_1,g^-_2,g^+_3,\gamma^+_4)
  =4\,g^{2}\langle 12\rangle^4 \sum_{\sigma } { {\rm Tr} \, (\,
    T^{a_{1_{\sigma}}}T^{a_{2_{\sigma}}}T^{a_{3_{\sigma}}}T^{a_{4}})\
    V(k_{1_{\sigma}},k_{2_{\sigma}},k_{3_{\sigma}},k_{4})\over\langle
    1_{\sigma}2_{\sigma} \rangle\langle
    2_{\sigma}3_{\sigma}\rangle\langle 3_{\sigma}4\rangle \langle
    41_{\sigma}\rangle }\ ,
\end{equation}
where the sum runs over all 6 permutations $\sigma$ of $\{1,2,3\}$ and
$i_{\sigma}\equiv\sigma(i)$. Note that in the effective field theory
of gauge bosons there are no Yang-Mills interactions that could
generate this scattering process at the tree level. Indeed, $V=1$ at
the  leading order of Eq.(\ref{vexp}) and the amplitude vanishes
due to the following identity:
\begin{equation}\label{ymlimit}
{1\over\langle 12\rangle\langle
      23\rangle\langle 34\rangle\langle
      41\rangle}+{1\over\langle 23\rangle\langle
      31\rangle\langle 14\rangle\langle 42\rangle}+{1\over\langle 31\rangle\langle
      12\rangle\langle 24\rangle\langle 43\rangle} ~=~0\ .
\end{equation}
Similarly,
the antisymmetric part of
the color factor (\ref{colf}) cancels out in the full amplitude (\ref{afull}). As a result,
one obtains:
\begin{equation}\label{mhva}
{\cal
    M}(g^-_1,g^-_2,g^+_3,\gamma^+_4)=8\, Q_c\, d^{abc}g^{2}\langle
  12\rangle^4\left({\mu(s,t,u)\over\langle 12\rangle\langle
      23\rangle\langle 34\rangle\langle
      41\rangle}+{\mu(s,u,t)\over\langle 12\rangle\langle
      24\rangle\langle 13\rangle\langle 34\rangle}\right),
\end{equation}
 where
\begin{equation}
\label{mudef}
\mu(s,t,u)= \Gamma(1-u)\left( {\Gamma(1-s)\over
      \Gamma(1+t)}-{\Gamma(1-t)\over \Gamma(1+s)}\right) .
\end{equation}
All
non-vanishing amplitudes can be obtained in a similar way. In
particular,
\begin{equation}
\label{mhvb}
{\cal M}(g^-_1,g^+_2,g^-_3,\gamma^+_4)=8\, Q_c\,
  d^{abc}g^{2}\langle 13\rangle^4\left({\mu(t,s,u)\over\langle
      13\rangle\langle 24\rangle\langle 14\rangle\langle
      23\rangle}+{\mu(t,u,s)\over\langle 13\rangle\langle
      24\rangle\langle 12\rangle\langle 34\rangle}\right),
\end{equation}
and the remaining ones can be obtained either by appropriate
permutations or by complex conjugation.

In order to obtain the cross section for the (unpolarized) partonic
subprocess $gg\to g\gamma$, we take the squared moduli of individual
amplitudes, sum over final polarizations and colors, and average over
initial polarizations and colors. As an example, the modulus square of
the amplitude (\ref{afull}) is:
\begin{equation}
\label{mhvsq}
|{\cal
    M}(g^-_1,g^-_2,g^+_3,\gamma^+_4)|^2=64\, Q_c^2\, d^{abc}d^{abc}g^{4}
  \left|{s\mu(s,t,u)\over u}+{s\mu(s,u,t)\over t} \right|^2 \, .
\end{equation}
 Taking
into account all $4(N^2-1)^2$ possible initial polarization/color
configurations and the formula~\cite{groupf}
\begin{equation}
\label{dsq}
\sum_{a,b,c}d^{abc}d^{abc}={(N^2-1)(N^2-4)\over 16 N},
\end{equation}
 we
obtain the average squared amplitude~\cite{Anchordoqui:2007da}
\begin{equation}
\label{mhvav}
|{\cal M}(gg\to
  g\gamma)|^2=g^4Q_c^2C(N)\left\{ \left|{s\mu(s,t,u)\over
        u}+{s\mu(s,u,t)\over t} \right|^2+(s\leftrightarrow
    t)+(s\leftrightarrow u)\right\},
\end{equation}
 where
\begin{equation}\label{cnn}
C(N)={2(N^2-4)\over N(N^2-1)}.
\end{equation}

Next, we consider the amplitude involving two $SU(N)$ gluons
$g_1,~g_2$ and two $U(1)$ gauge bosons $\gamma_3,\gamma_4$ associated to
the same $U(N)$ quiver:
\begin{equation}
\label{gens2}
T^{a_1}=T^a \ ,~ \ T^{a_2}=T^b\ ,~ \
  T^{a_3}=Q_cI \ ,~ \ T^{a_4}=Q_cI\ .
\end{equation}
This amplitude can be obtained from $gg\to g\gamma$ by replacing
$d^{abc}$ with ${1\over 2}Q_c\delta^{ab}$. Hence at the level of squared
amplitudes, summed over final polarizations and colors and averaged
over initial polarizations and colors~\cite{photons}
\begin{equation}
\label{mhvgg}|{\cal M}(gg\to
  \gamma\gamma)|^2={4N Q_c^2 \over N^2-4}\,
|{\cal M}(gg\to
  g\gamma)|^2.
  \end{equation}

The two most interesting energy regimes of $gg\to g\gamma$
scattering are far below the string mass scale $M_s$
and near the threshold for the production of massive string
excitations. At low energies, Eq.~(\ref{mhvav}) becomes
\begin{equation}
\label{mhvlow}
|{\cal M}(gg\to
  g\gamma)|^2\approx g^4Q_c^2C(N){\pi^4\over 4}(s^4+t^4+u^4)\qquad
  (s,t,u\ll 1) \, .
\end{equation}
The absence of massless poles, at $s=0$ {\it etc.\/}, translated
into the terms of effective field theory, confirms that there are
no exchanges of massless particles contributing to this process.
On the other hand, near the string threshold $s\approx M_s^2$
(where we now restore the string scale)
\begin{equation}
\label{mhvlow2}
|{\cal M}(gg\to g\gamma)|^2\approx
4g^4Q_c^2C(N){M_s^8+t^4+u^4\over M_s^4[(s-M_s^2)^2+(\Gamma M_s)^2]}
\qquad (s\approx M_s^2),
\end{equation}
with the singularity (smeared with a width $\Gamma$) reflecting the
presence of a massive string mode propagating in the $s$
channel. (Further details are given in Appendix C.) It
should be noted that because the $g \gamma$ final state projects onto
pure color octet, only the $SU(3)$ adjoint string excitations ($G^*$)
contribute to Eq.~(\ref{mhvlow2})~\cite{widths}. On the other hand, for $gg \to
\gamma \gamma$ only the color singlet excitation ($C^{0*}$) is present in
the intermediate state.

An important modification needs to be introduced into
Eq.~(\ref{mhvlow2}), because it contains additively contributions from
both angular momenta $J=0$ and $J=2$, corresponding to incoming
helicities ($\pm \pm$) and ($\pm \mp$), respectively. In general these
contributions would have different widths, for the $G^*$ excitation:
$\Gamma^{J=0} = 75 \, (M_s/{\rm TeV})~{\rm GeV}$ and $\Gamma^{J=2} =
45 \, (M_s/{\rm TeV})~{\rm GeV}$~\cite{widths}.  These widths are
premised on the assumption that corrections of order $(M_{Z'}/M_s)^2$
are negligible, both in obtaining matrix elements and in calculating
phase space.

In what follows we will take $N=3$ and set $g$ equal to the QCD coupling
constant, $\alpha_s = (g^2/4\pi) \sim 0.1$. Before
proceeding with numerical calculation, we need to make precise the
value of $Q_c$. If we were considering the process $gg\rightarrow C^0 g,$
where $C^0$ is the $U(1)$ gauge field tied to the $U(3)$ brane, then $Q_c =
\sqrt{1/6}$ due to the normalization condition~(\ref{norm}). However,
for $gg\rightarrow \gamma g$ there are two additional projections:
from $C_\mu$ to the hypercharge boson $Y_\mu$, giving a mixing factor
$\kappa$; and from $Y_\mu$ onto a photon, providing an additional
factor $\cos\theta_W \ (\theta_W=$ Weinberg angle). The $C^0-Y$ mixing
coefficient is model dependent: in the minimal
model~\cite{Berenstein:2006pk} it is quite small, around $\kappa
\simeq 0.12$ for couplings evaluated at the $Z$ mass, which is
modestly enhanced to $\kappa \simeq 0.14$ as a result of RG running of
the couplings up to 2.5~TeV.  It should be noted that in
models~\cite{ant,bo} possessing an additional $U(1)$ which partners
$SU(2)_L$ on a $U(2)$ brane, the various assignment of the charges can
result in values of $\kappa$ which can differ considerably from
$0.12.$ In what follows, we take as a fiducial value $\kappa^2 =
0.02.$ Thus, if (\ref{mhvlow2}) is to describe $gg\rightarrow \gamma
g,$ we modify our definition of $Q_c$ given in Eq.~(\ref{gens}) to
accommodate the additional mixings, and obtain
\begin{equation}
Q_c^2= \tfrac{1}{6} \ \kappa^2 \ \cos^2\theta_W \simeq 2.55\times
10^{-3}\ \left(\kappa^2/0.02\right)\ \ .
\label{Q2}
\end{equation}
In the remainder of this chapter, we explore potential searches for
 Regge excitations of fundamental strings at the LHC.

\section{High-${k}_{\perp} $ Isolated Photons}

In order to assess the possibility of discovery of signal above QCD
background, we adopt the kind of signal introduced
in~\cite{Dimopoulos:2001hw} to study detection of TeV-scale black
holes at the LHC, namely a high-$k_{\perp}$ isolated $\gamma$ or $Z.$
Thus, armed with parton distribution functions
(CTEQ6D)~\cite{Pumplin:2002vw,Stump:2003yu}, in what follows we
calculate integrated cross sections $\sigma(pp\rightarrow \gamma +
{\rm jet})|_{k_{\perp }(\gamma)> k_{\perp, {\rm min}}}$ for both the
background QCD processes and for $gg\rightarrow \gamma g$, for an
array of values for the string scale $M_s$.

\subsection{QCD background}

\begin{figure}[tbp]
\postscript{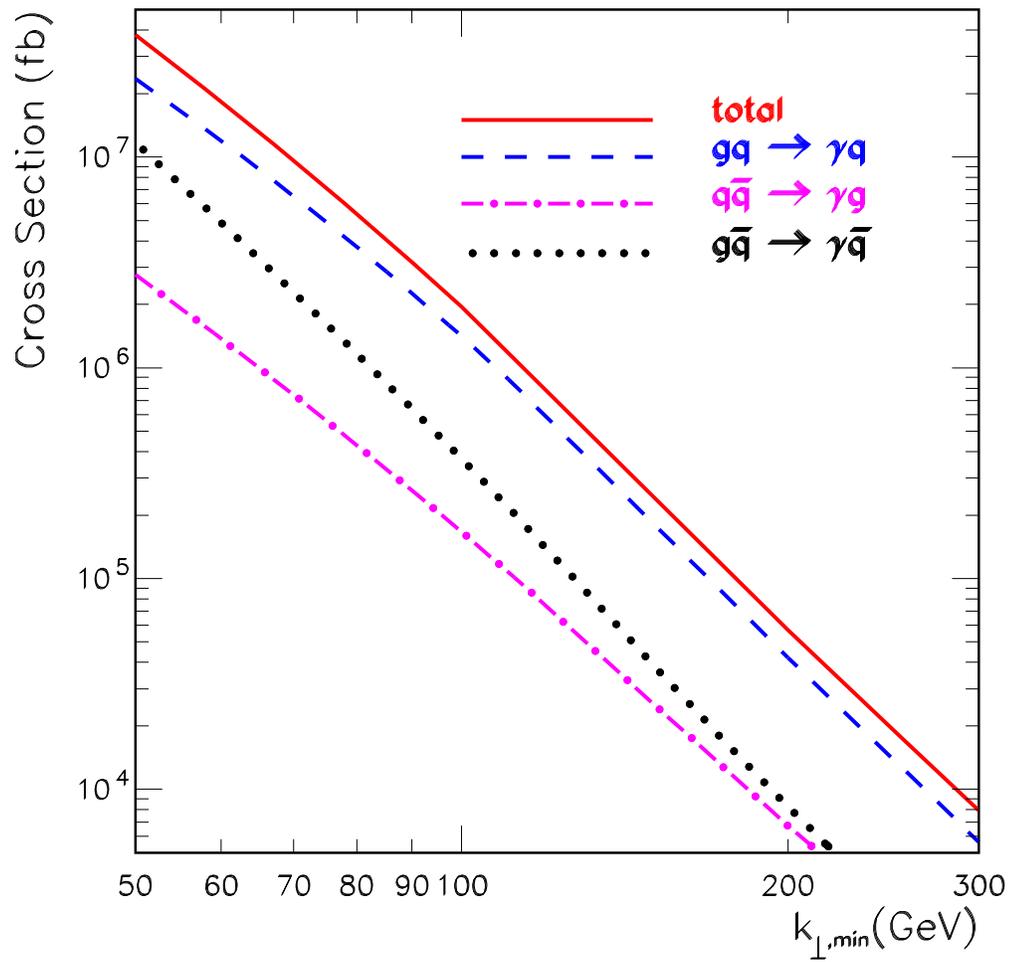}{0.9}
\caption[Cross section of QCD background]{Different contributions to the QCD cross section for $pp \to
  \gamma + {\rm jet}$ as a function of $k_{\perp, {\rm min}}$. It is
  clearly seen that the $gq \to \gamma q$ process provides the
  dominant contribution.}
\label{fig:qcd_B}
\end{figure}

The SM background for processes with a single photon in the final
state originates in the parton tree level processes $g q \rightarrow
\gamma q,\ g\bar q\rightarrow \gamma\bar q\ {\rm and} \ q\bar
q\rightarrow \gamma g$,
\begin{eqnarray}
\left. 2 E' \frac{d\sigma}{d^3k'} \right|_{pp \to \gamma X} & = &  \sum_{ijk} 
\left. \int dx_a \, dx_b \, f_i(x_a,Q) \, f_j (x_b,Q) \, 2 E' \frac{d \hat \sigma}{d^3k'}\right|_{ij \to \gamma k} \ ,
\end{eqnarray} 
where $x_a$ and $x_b$ are the longitudinal fractions of momenta of the parent
hadrons carried by the partons which collide, $k'$ $(E')$ is the
photon momentum (energy), $d \hat \sigma/d^3k'|_{ij \to \gamma k}$ is
the cross section for scattering of partons of type $i$ and $j$
according to elementary QCD diagrams, $f_i(x_a,Q)$ and $f_j (x_b, Q)$
are parton distribution functions, $Q$ is the momentum transfer, and
the sum is over the parton species: $g, q = u,\ d,\ s,\
c,\ b$. In what follows, we focus on $gq \to \gamma q$, which results
in the dominant contribution to the total cross section. Corrections
from the other two processes can be computed in a similar fashion.
The hard parton-level cross section reads,
\begin{equation} 
  2 E' \frac{d \hat \sigma}{d^3k'} (g(k)\,q(p) \to \gamma(k')\, q(p))  =  \frac{1}{(2 \pi)^2} \, \frac{1}{2 \hat s} \, \delta[(k+p-k')^2] 
  \, \frac{1}{4}\sum |{\cal M}|^2 
\label{under}
\end{equation}
where the variables $k$, $p$, $k'$ and $p'$ in the parentheses are the momenta of the partons. Here, the amplitude for $gq \to \gamma q$ is given by
\begin{equation}
 \frac{1}{4} \sum |{\cal M}|^2\, = \frac{1}{3} g^2 e^2 e_q^2 \left(\frac{\hat s}{\hat s + \hat t} + \frac{\hat s + \hat t}{\hat s} \right),
\end{equation}
where $\hat s=(k+p)^2$, $ \hat t=(k-k')^2$ and $\hat u=(k-p')^2$ are the Mandelstam variables in the parton level, $g$ and $e$ are the QCD and electromagnetic coupling constants,
and $e_q$ is the fractional electric charge of species $q$. For
completness we note that for $q \bar q \to g \gamma$,
\begin{equation}
 \frac{1}{4} \sum |{\cal M}|^2 \, = \frac{8}{9} g^2 e^2 e_q^2 \left(- 
\frac{\hat t}{\hat s + \hat t} - \frac{\hat s + \hat t}{\hat t} \right)\, .
\end{equation}
In this process, we assume that the proton momenta take the explicit forms in the $pp$ center of mass frame
\begin{equation}\label{protonmomenta}
P_1 = (\sqrt{s}/2,\, 0,\, 0,\, \sqrt{s}/2),  \ \ \ \ \ \ P_2 = (\sqrt{s}/2,\, 0,\, 0,\, -\sqrt{s}/2)
\end{equation}
and the final photon has momentum
\begin{equation}\label{rapidity}
k'_0 = k_\perp \, \cosh y, \ \ \ \ \ \  k'_\parallel = k_\perp \, \sinh y \, ,
\end{equation}
where $k_\perp$ is the transverse momentum of the photon and $y$ is called the longitudinal rapidity. The relation between the momenta of protons $P_1$, $P_2$ and those of the incoming partons $k$, $p$ can be written with the longitudinal fractions $x_a$, $x_b$:
\begin{equation}\label{longitudinal}
k = x_a P_1, \ \ \ \ \ \ p = x_b P_2 \, .
\end{equation}
Using Eqs. (\ref{protonmomenta}), (\ref{rapidity}) and (\ref{longitudinal}) , we can re-write the argument of the delta function as
\begin{eqnarray}
(k+p-k')^2&=& 2\, x_b \, P_2 \ . \ (x_a P_1 - k') + \hat t \nonumber \\
&=& x_a\, x_b\, s - 2\, x_b \, P_2 \ . \ k' + \hat t \nonumber \\
&=& x_a\, x_b\, s - \sqrt{s}\, x_b\, k_\perp \, e^y - \sqrt{s} \, x_a \, k_\perp \, e^{-y} \, .
\end{eqnarray}
so that
\begin{eqnarray}
 \delta[(k+p-k')^2] &=&\delta (x_a\, x_b\, s - \sqrt{s}\, x_b\, k_\perp \, e^y - \sqrt{s} \, x_a \, k_\perp \, e^{-y})  \nonumber \\
 & = & \frac{1}{s \, \left[x_a - x_\perp \, e^{y}\right]} \, \, \, \delta \left(x_b - \frac{x_a \, x_\perp \, e^{-y}}{x_a - x_\perp \, e^y}\right), 
\end{eqnarray}
where $x_\perp = k_\perp /\sqrt{s}.$ The lower bound $x_b > 0$ implies $x_a > x_\perp \, e^y$. The upper bound $x_b < 1$ leads to a stronger constraint  
\begin{equation}
x_a > \frac{x_\perp e^y}{1 - x_\perp e^{-y}} \,,
\label{bound}
\end{equation}
which requires $x_\perp e^y < 1 - x_\perp e^{-y}$, yielding $x_\perp <
(2 \, {\rm cosh}\, y)^{-1}$. Of course there is another completely
symmetric term, in which $g$ comes from $P_2$ and $q$ comes from
$P_1$.  Putting all this together, the total contribution from $gq \to
\gamma q$ reads
\begin{eqnarray}
  \sigma_{pp \to \gamma X}^{qg \to \gamma q} & = &  2\, \sum_q \int \frac{d^3k'}{2E'} \int dx_a \int dx_b \, f_g(x_a,Q) \, f_{q}(x_b,Q) \, \frac{1}{(2\pi)^2} \, \, 
 \frac{1}{s \, \left[x_a - x_\perp e^y \right]} \nonumber \\ 
 & \times &  \frac{1}{2 \hat s} \, \, \delta \left(x_b - \frac{x_a x_\perp e^{-y}}{x_a - x_\perp e^{y}} \right) \, \, \frac{e^2 g^2 e_q^2}{3} \, \, \left(\frac{\hat s + \hat t}{\hat s} + \frac{\hat s}{\hat s + \hat t} \right) \, .
\label{laven}
\end{eqnarray} 
With the change of variables $z = e^y$ and the relation
\begin{equation}
\frac{d^3k'}{2E'} = \pi k_\perp\, 
dk_\perp\, dy\, = \frac{\pi \, k_\perp \, dk_\perp \, dz}{z} \, , \ \ \ \ \ \ \frac{\hat t}{\hat s} = - \frac{\sqrt{s} k_\perp e^{-y}}{x_b s} = -\frac{x_\perp}{x_b \, z} \, ,
\end{equation}
 Eq.~(\ref{laven}) can be re-written as
\begin{eqnarray}
 \sigma_{pp \to \gamma X}^{qg \to \gamma q} & = & \frac{e^2 g^2}{12 \pi s}\, \int_{x_{\perp {\rm min}}}^{1/2} dx_\perp \, \int_{z_{\rm min}}^{z_{\rm max}}  dz \int_{x_{a,min}}^1 dx_a\,\, f_g(x_a,Q) \left[\sum_q e_q^2\,\, f_{q}\left( \frac{ x_a x_\perp z^{-1}}{x_a - x_\perp z},Q \right) 
\right] \nonumber \\
 & \times & \frac{1}{x_a^2} \, \left(\frac{x_\perp z}{x_a} + \frac{x_a}{x_\perp z}\right) \,,
\label{hc}
\end{eqnarray}
where the integration limits,
\begin{equation}
z_{^{\max}_{\rm min}} = \frac{1}{2} \left[ \frac{1}{x_\perp} \pm \sqrt{\frac{1}{x_\perp^2} -4} \right] \ \ \ \ \ \ \ \ \ \ \ \ {\rm and} \ \ \ \ \ \ \ \ \ \ \ \   
x_{a,{\rm min}} = \frac{x_\perp z}{1 - x_\perp z^{-1}} \, ,
\end{equation}
are obtained from Eq.~(\ref{bound}). In Fig.~\ref{fig:qcd_B} we show
the QCD background cross section {\em vs} $k_{\perp, {\rm min}}$, as
obtained through numerical integration of Eq.~(\ref{hc}).  To
accommodate the minimal acceptance cuts on final state photons from the
CMS~\cite{Ball:2007zza} and ATLAS~\cite{Armstrong:1994it} proposals, an
additional kinematic cut, $|y|<2.4,$ has been included in the
calculation.

\subsection{The string signal}

\begin{figure}
 \postscript{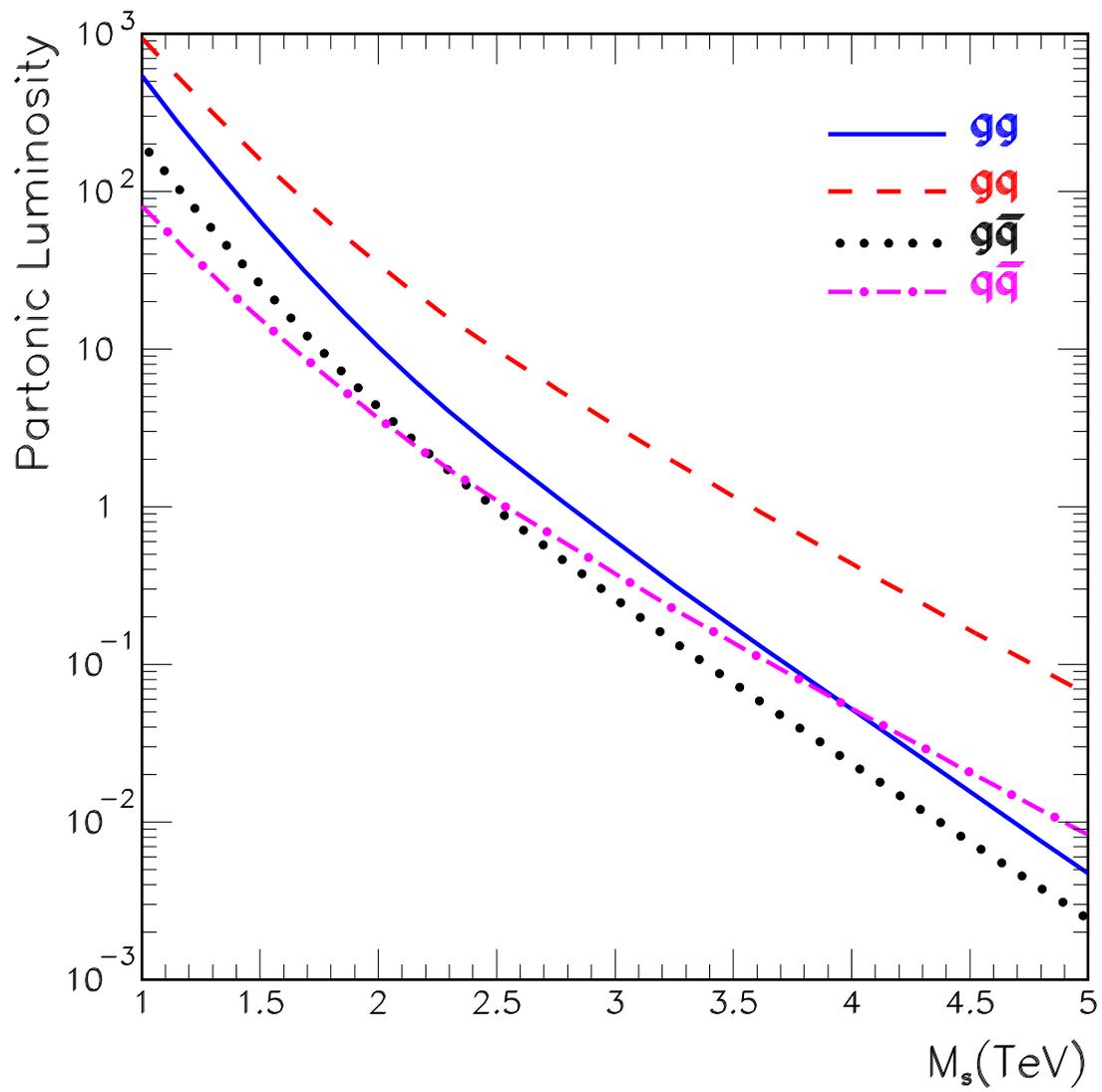}{0.98}
 \caption[Partonic luminosity]{Relative contributions of initial state partons ($ij = gg,\ gq,\ g\bar q,\ {\rm and}\ q \bar q$) to $\int_{\tau_0}^1 f_i(x_a, Q) \, \, f_j(\tau_0/x_a, Q)\,\,\, dx_a/x_a$, with varying string scale.}
\label{fig:pl}
\end{figure}

For the considerations in this Dissertation, the resonant cross section
can be safely approximated by single poles in the Narrow-Width
Approximation,
\begin{equation}
\frac{\Gamma \sqrt{s_0}/\pi}{(\hat s - s_0)^2 + (\Gamma \sqrt{s_0})^2}\, \frac{\pi}{\Gamma \sqrt{s_0}} =  \frac{\pi}{\Gamma \sqrt{s_0}} \,\, \delta(\hat s - s_0) \,,
\end{equation}
where $s_0 = M_s^2$.  The scattering proceeds through $J=0$ and $J=2$
angular momentum states, with the $M_s^8$ term in Eq.~(\ref{mhvlow2})
originating from $J=0$, and the $t^4+ u^4$ piece reflecting $J=2$
activity. The widths of these two resonances are different, with
$\Gamma^{J=0} = (3/4) \, \alpha_s M_s,$ and $\Gamma^{J=2}= (9/20)\,
\alpha_s M_s$~\cite{widths}.  The average string amplitude square in
Eq.~(\ref{mhvlow2}) then becomes
\begin{eqnarray}
|{\cal M}(gg\to g\gamma)|^2 & \approx &
4g^4Q_c^2C(N) \, \frac{\pi}{s_0^{5/2}}
\, \left[\frac{s_0^4}{\Gamma^{J=0}}+ \frac{\hat t^4+ (\hat t + s_0)^4}{
\Gamma^{J=2}} 
\right] \, \delta(\hat s - s_0) \nonumber \\
 & = & 4g^4Q_c^2C(N) \, \frac{\pi}{\alpha_s \, s_0^{3}}
\, \left\{\tfrac{4}{3} s_0^4+ \tfrac{20}{9} [\hat t^4+ (\hat t + s_0)^4] 
\right\} \, \delta(\hat s - s_0) \, . 
\end{eqnarray}
Thus, the total cross section for single photon
production in gluon fusion is given by
\begin{eqnarray}
  \sigma_{pp \to \gamma X}^{gg \to \gamma g} & = &  \int \frac{d^3k'}{2E'} \int dx_a \int dx_b \, f_g(x_a,Q) \, f_{g}(x_b,Q) \, \frac{1}{(2\pi)^2} \, \,
 \frac{1}{2\,\hat s\, s} \delta(x_a\, x_b -x_b x_\perp z - x_a x_\perp z^{-1}) \nonumber \\
 & \times & 4 g^4 Q_c^2 C(N)  \frac{\pi}{\alpha_s \, s_0^3} \, \left\{ 
\tfrac{4}{3} s_0^4 + \tfrac{20}{9} [\hat t^4 + (\hat t + s_0)^4 ] \right\}\, \delta (\hat s - s_0) \,.
\label{mel}
\end{eqnarray}
We set $Q = M_s$, which is appropriate for the dual picture of string
theory. We are aware that for $Q \sim M_s$, the parton distribution
functions will receive significant corrections from the rapid increase
of degrees of freedom. Fortunately, as noted
elsewhere~\cite{Anchordoqui:2001cg}, at parton center-of-mass energies
corresponding to low-lying string excitations the resonant cross
section is largely insensitive to the details of the choice of $Q$.
Since the second delta function can be written as
\begin{equation}
\delta (\hat s - s_0) =\frac{1}{x_a s} \delta (x_b -\frac{s_0}{x_as}) \, ,
\end{equation}
integration over $x_b$ leads to
\begin{eqnarray}
  \sigma_{pp \to \gamma X}^{gg \to \gamma g} & = & \frac{g^4 Q_c^2 C(N)}{2\, \alpha_s \tau_0^4 s} \int \frac{x_\perp \, dx_\perp \, dz}{z} \int dx_a \, f_g(x_a,Q) \, f_{g}(\tau_0/x_a,Q) \,\frac{1}{x_a} \nonumber \\
 & \times &\,\,
 \delta \left(\tau_0 - \frac{\tau_0 x_\perp z}{x_a} - \frac{x_a x_\perp}{ z}\right)   \, \left\{ \tfrac{4}{3} \tau_0^4 + \tfrac{20}{9} [ (x_a \, x_\perp z^{-1})^4 + (-x_a\, x_\perp \, z^{-1} + \tau_0)^4] \right\} , \nonumber \\
 & &
\end{eqnarray}
where $\tau_0 = s_0/s$. In order to proceed the integral over the argument $z$, we write
\begin{equation}
f(x)\equiv \tau_0 - \frac{\tau_0 x_\perp z}{x_a} - \frac{x_a x_\perp}{ z} \, .
\end{equation}
Then, the delta function becomes of the form:
\begin{equation}
 \delta(f(z)) = \frac{1}{|f'(z_+)|} \, \, \delta (z - z_+) + \frac{1}{|f'(z_-)|} \, \, \delta (z - z_-) \,,
\end{equation}
where $z_\pm$ are the solutions to $f(z) = 0$,
\begin{equation}
z_\pm = \frac{x_a}{2 x_\perp}\, \left( 1 \pm \sqrt{1 - \frac{4 x_\perp^2}{\tau_0}} \right) \, \,.
\end{equation}
Using the identities
\begin{equation}
\frac{1}{z_\pm |f'(z_\pm)| }= \left| \frac{\tau_0 x_\perp z_\pm}{x_a} - \frac{x_a \, x_\perp}{z_\pm} \right|^{-1}
\end{equation} 
and
\begin{eqnarray}
\tfrac{16}{9} \, \tau_0^2 \, (5 \, x_\perp^4  -10 \, x_\perp^2\, \tau_0  
+ 4 \, \tau_0^2) 
& = &
\left\{\tfrac{4}{3}\, \tau_0^4 + \tfrac{20}{9} \, [(x_a x_\perp z_+^{-1})^4 + (-x_a x_\perp z_+^{-1} + \tau_0)^4] \right\} \nonumber \\ & + &
\left\{ \tfrac{4}{3} \tau_0^4 + \tfrac{20}{9} \, [ (x_a x_\perp z_-^{-1})^4 + (-x_a x_\perp z_-^{-1} + \tau_0)^4] \right\} \,, \nonumber \\
 & &
\end{eqnarray}
the integral over the $z$ variable yields
\begin{eqnarray}
  \sigma_{pp \to \gamma X}^{gg \to \gamma g} & = & \frac{8}{9} \, \frac{g^4 \, Q_c^2 C(N)}{\alpha_s \, \tau_0^3 \, s} \, \int_{x_{\perp, {\rm min}}}^{\sqrt{\tau_0}/2} d x_\perp \, \frac{x_\perp}{\sqrt{1- 4x_\perp^2/\tau_0}} \,\, \left(5 \,  x_\perp^4 - 10 \, x_\perp^2 \, \tau_0 + 4 \, \tau_0^2 \right)  \nonumber \\
 & \times & \int_{\tau_0}^1 \frac{dx_a}{x_a} \, \, f_g(x_a,Q) \, \, f_g (\tau_0/x_a,Q)  \, ,
\end{eqnarray}
where the integration range has been derived from the conditions
$0 < x_b = \tau_0/x_a < 1$ and $4 x_\perp^2<\tau_0$, which imply $\tau_0 < x_a < 1$ and
$x_{\perp, {\rm min}} < x_\perp < \sqrt{\tau_0}/2$.
Finally, integration over $x_\perp$ leads to
\begin{eqnarray}
  \sigma_{pp \to \gamma X}^{gg \to \gamma g} & = & \frac{1}{9} \frac{g^4\, Q_c^2 C(N)}{\alpha_s \ \tau_0^2 \, s} \, \sqrt{1 - \frac{4 x_{\perp, {\rm min}}^2}{\tau_0}} \left(5\  \tau_0^2 - 6\  \tau_0 \ x_{\perp, {\rm min}}^2 + 2 \ x_{\perp, {\rm min}}^4\right) \nonumber \\
 & \times & \int_{\tau_0}^1 \frac{dx_a}{x_a} \, \,  f_g(x_a,Q) \, \, f_g (\tau_0/x_a,Q)  \, .
\label{csb}
\end{eqnarray}
\begin{figure}
\postscript{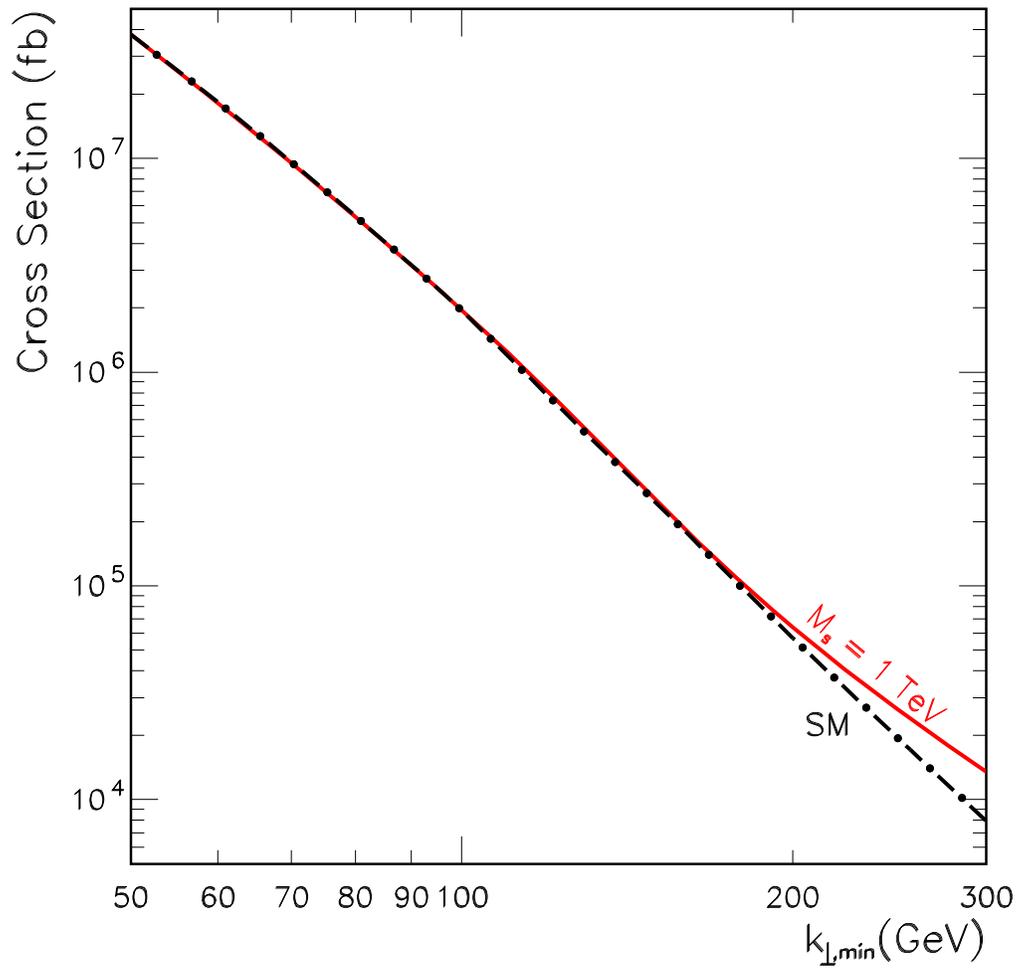}{0.98}
\caption[Resonant cross section with varying $k_{\perp,{\rm min}}$]  
{Behavior of the QCD cross section for $pp \to \gamma + {\rm
    jet}$ (dot-dashed line) as a function of $k_{\perp, {\rm min}}$.
  The string cross section overlying the QCD background is also shown
  as a solid line, for $M_s = 1$~TeV.}
\label{fig:ptfig}
\end{figure}
In Fig.~\ref{fig:ptfig} we show the resonant cross section 
for $M_s = 1$~TeV. It
is evident that the background is significantly reduced for large
$k_{\perp, {\rm min}}$. At very large values of $k_{\perp, {\rm min}},$ however,
event rates become problematic. Note that all stringy corrections to the pure bosonic cross section
given by Eq.~(\ref{csb}) have similar factorizations. An illustration of the
relative partonic luminosities of the different processes is shown in
Fig.~\ref{fig:pl}.

\subsection{LHC discovery reach}

\begin{figure}
\postscript{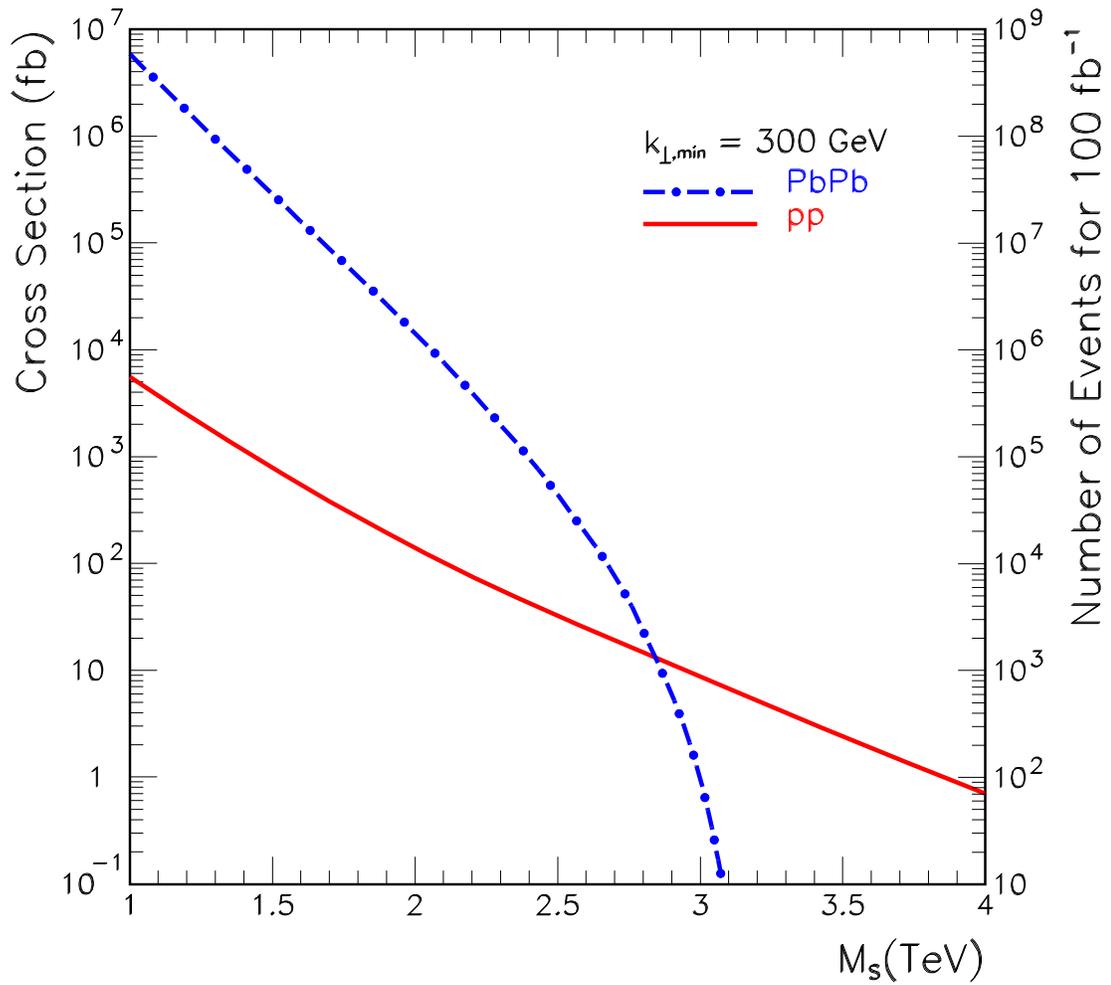}{0.98}
\caption[Resonant cross section with varying string scale]{Cross section for gluon fusion into $\gamma + {\rm jet}|_{p_{\rm
      T}(\gamma)> 300~{\rm GeV}}$ and expected number of
  events, for $100~{\rm fb}^{-1}$ and varying string scale~\cite{photons}.}
\label{fig:sigma}
\end{figure}

\begin{figure}
 \postscript{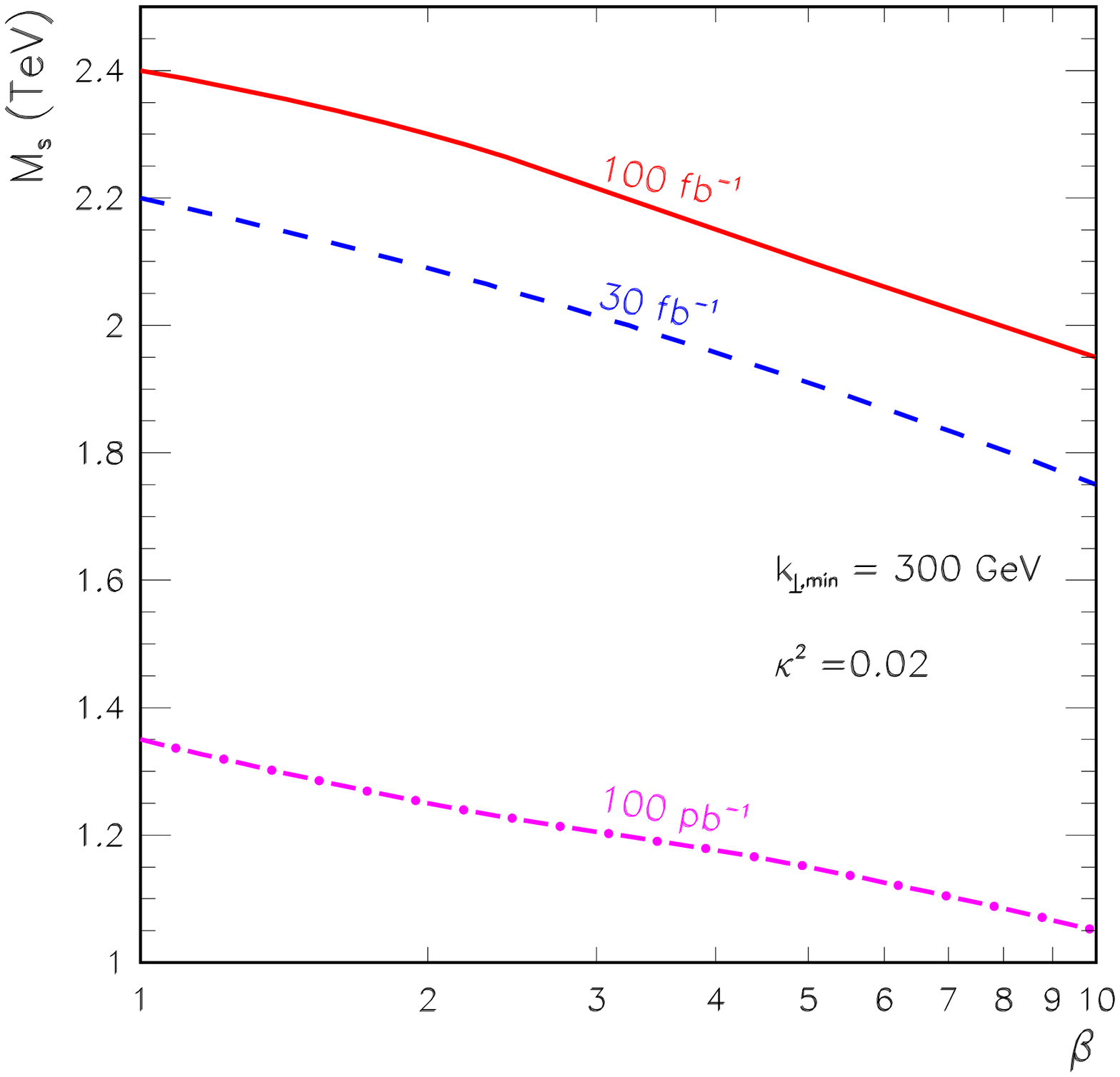}{0.98}
 \caption[LHC discovery reach]{Contours of 5$\sigma$ discovery in the ($M_s$, detector
   efficiency) plane for different integrated luminosities and
   $\kappa^2 =0.02$.}
\label{fig:S2N}
\end{figure}

In this section we explore the LHC discovery potential
by computing the signal-to-noise ratio (${\rm signal}/\sqrt{\rm SM\
  background}$). In Fig.~\ref{fig:sigma} we show the
string cross section and number of events (before cuts) in a 100
fb$^{-1}$ run at the LHC, for $p_{\rm T,min}=300$ GeV, as a function of
the string scale $M_s$.  For a 300~GeV cut in the transverse momentum, the QCD
cross section (shown in Fig.~\ref{fig:ptfig}) is about $8 \times
10^3$~fb, yielding (for 100~fb$^{-1}$) $\sqrt{\rm SM\ background}
\approx 895.$  A point worth noting at this juncture: In order to
minimize misidentification with a high-$k_\perp \ \pi^0$, isolation
cuts must be imposed on the photon, and to trigger on the desired
channel, the hadronic jet must be identified~\cite{Bandurin:2003kb}.
We will leave the exact nature of these cuts for the experimental
groups, and present results for a generous range of direct photon
reconstruction efficiency. To do so, we define the parameter
\begin{equation}            
 \beta = \frac{{\rm background \, due \, to \, misidentified} \, \pi^0 \, {\rm 
after \, isolation \, cuts}}{{\rm QCD \, background \, from \, direct \, photon \, production}}  + 1 \,\, .   
\end{equation}
Therefore, the noise is increased by a factor of $\sqrt{\beta}$, over
the direct photon QCD contribution.  Our significant results are
encapsuled in Fig.~\ref{fig:S2N}, where we show the discovery reaches
of the LHC for several integrated luminosities and
$\kappa^2 = 0.02$.  A detailed study of the CMS potential for
isolation of prompt-$\gamma$'s has been recently carried
out~\cite{Gupta:2007cy}, using GEANT4 simulations of $\gamma + {\rm
  jet}$ events generated with Pythia. This analysis (which also
includes $\gamma$'s produced in the decays of $\eta$, $K_s^0$,
$\omega^0,$ and bremsstrahlung photons emerging from high-$p_\perp$
jets) suggests $\beta \simeq 2$. Of course, considerations
of detector efficiency further reduce the $S/N$ ratio by an additional
factor $\epsilon$, where $1 < \epsilon \ll \sqrt{\beta}$. We conclude
that {\em discovery at the LHC would be possible for $M_s$ as large as
  2.3~TeV.} 

We now consider gluon fusion into two photons.  As can be seen in
Eq.~(\ref{mhvgg}), the string amplitude for diphoton production is
suppressed by a factor of $Q_c^4$. On the other hand, the QCD leading
order contribution to diphoton production, given by the Born level
process $q\bar q \to \gamma \gamma,$ appears with an additional
$\alpha_{em}$ so that it nearly compensates the extra factor
$\kappa^2$. However, the restriction to $C^{0*}$ of the intermediate
state, introduces an important dependence on the unknown $C^{0}$ mass,
because the pole is shifted away from $M_s$. Additionally, the widths
of the $C^{0*}$ excitation, $\Gamma^{J=0} = 150 \, \alpha_s \, (M_s/{\rm
  TeV})~{\rm GeV} \ {\rm and} \ \Gamma^{J=2} = 75 \, \alpha_s \,
(M_s/{\rm TeV})~{\rm GeV}$~\cite{widths}, are nearly twice that of the
$G^*$. These two considerations vitiate any useful sensitivity of the
diphoton channel.

We now briefly explore the potential of the ALICE to search for low mass
string excitations.\footnote{Pb-Pb $\to \gamma$ + jet events can be
  identified by selecting a prompt photon and searching for the
  leading particle in the opposite direction inside the ALICE central
  tracking system~\cite{Conesa:2007nx}.  As photons emerge almost
  unaltered from dense medium, they provide a measurement of the
  original energy of the parton emitted in the opposite direction.}
With this motivation, we extend our analysis to include heavy ions
collisions. In the spirit of Ref.~\cite{Chamblin:2002ad} we consider
the unshadowed parton distribution functions, i.e.,
\begin{equation}
R_{i/A} (x) = \frac{f_{i/A} (x,Q)}{A f_{i} (x,Q)}
\simeq 1 \, ,
\end{equation}
where $f_{i/A}$ and $f_{i}$ are the parton distribution functions
inside a free nucleus of mass $A$ and free nucleon, respectively.  For
$M_s > 1$~TeV, this approximation holds because the LHC Pb-Pb collisions
probe the minimum value of parton momentum at $x_{\rm min} \approx
M_s^2/s \sim 0.033,$ where there are no shadowing effects. A
comparison of the string cross section for gluon fusion into $\gamma +
{\rm jet}|_{k_{\perp}(\gamma)> 300~{\rm GeV}}$ for $pp$ and Pb-Pb
collisions is shown in Fig.~\ref{fig:sigma}. However, the larger aggregate of
partons also increase the SM background; namely, for $k_{\perp, {\rm
    min}} > 300~{\rm GeV},$ $\sigma_{{\rm Pb-Pb} \to \gamma X} \approx
2.8 \times 10^7~{\rm fb}.$ This greatly decreases the sensitivity to D-brane 
models, which would 
require a Pb-Pb integrated luminosity of a few hundred pb$^{-1}$. This 
is substantially larger than the present day estimate~\cite{Dainese:2007dg}.

%% file: chap_bh.tex
If nature has gracefully picked a sufficiently low-scale gravity,
microscopic black holes can be produced at particle
accelerators~\cite{Banks:1999gd}. In particular, the cross section for
black hole production at the LHC is expected to be $\sim 100$~pb for a
fundamental Planck scale $M_{10} \sim 1$~TeV, which could turn the LHC
into a black hole factory with a production rate of $\sim
1$~Hz~\cite{Dimopoulos:2001hw,Giddings:2001bu}. The LHC Hawking
temperature would be few hundred GeV, and so black holes would quickly
evaporate into about a half dozen particles with large transverse
momentum. In this chapter we discuss potential methods to discriminate
the high-$k_\perp$ string decay products from the light descendants of
black holes. Interestingly, one of these methods allows an increase of
the LHC sensitivity for Regge recurrences of fundamental strings up to
about 4 TeV.

\section{Bump-Hunting}
\label{s4}

The discovery trigger described in the previous chapter, the
observation of isolated photons at large transverse momentum, serves
very well as a signature of new physics. However, as mentioned above, 
this criterion served also as a marker for Hawking
radiation following production of TeV-scale black holes at the LHC. Given
the particular nature of the string process we are considering, the
production of a TeV-scale resonance and its subsequent 2-body decay,
signatures in addition to large $k_\perp$ photons are available.
Most apparently, one would hope that the resonance would be visible in data
binned according to the invariant mass $M$ of the photon + jet,
setting cuts on photon and jet rapidities, $|y_1|,\, |y_2| < y_{\rm max}=2.4$,
respectively.  With the definitions $Y\equiv \thalf (y_1 + y_2)$ and
$y \equiv \thalf(y_1-y_2)$, the cross section per interval of $M$ for
$pp\rightarrow \gamma + {\rm jet} +X$ is given by~\cite{Eichten:1984eu}
\begin{eqnarray} \frac{d\sigma}{dM} & = &  \frac{M^3}{s}\ \sum_{ijk}\left[
\int_{-Y_{\rm max}}^{0} dY\,f_i (x_a,\, M)  \right. f_j (x_b, \,M ) \,
\int_{-(y_{\rm max} + Y)}^{y_{\rm max} + Y}\frac{dy}{\cosh^2
y}
\left. \frac{d\sigma}{d\hat t} \right|_{ij\rightarrow k,\gamma} \nonumber \\ 
& + &\int_{0}^{Y_{\rm max}} dY\, f_i (x_a, \, M)
f_j (x_b, M) \, \int_{-(y_{\rm max} - Y)}^{y_{\rm max} - Y} \frac{dy}{\cosh^2 y}
\left. \left. \frac{d\sigma}{d\hat t}\right|_{ij\rightarrow k,\gamma}  \right] 
\label{longBH}
\end{eqnarray}
where $i,j,k$ are different partons, and the longitudinal fractions have the 
forms
\begin{equation}
x_a =M e^{Y}/\sqrt{s}, \ \ \ \ \ \ x_b = M e^{-Y}/\sqrt{s} \, 
\label{longitudnal}
\end{equation}
(see Appendix D for details). The kinematics of
the scattering provides the relation
\begin{equation}
k_\perp = \frac{M}{2\, \cosh y} \, ,
\label{relation}
\end{equation}
which, when combined with the standard cut $k_\perp > k_{\perp,
  {\rm min}}$, imposes a {\em lower} bound on $y$ to be implemented in
the limits of integration.  The $Y$ integration range in
Eq.~(\ref{longBH}), $Y_{\rm max} = {\rm min} \{ \ln(\sqrt{s}/M),\ \
y_{\rm max}\}$, comes from requiring $x_a, \, x_b < 1$ together with the
rapidity cuts $|y_1|, \, |y_2| \le 2.4$. Finally, the Mandelstam
invariants occurring in the cross section are given by
\begin{eqnarray}
\hat s &= &M^2,  \nonumber \\
\hat t &=& - M^2 e^{-y}/2 \cosh y,  \nonumber \\
\hat u &=& - M^2 e^{+y}/2 \cosh y. \label{Mandelstam}
\end{eqnarray}
\begin{figure}
 \postscript{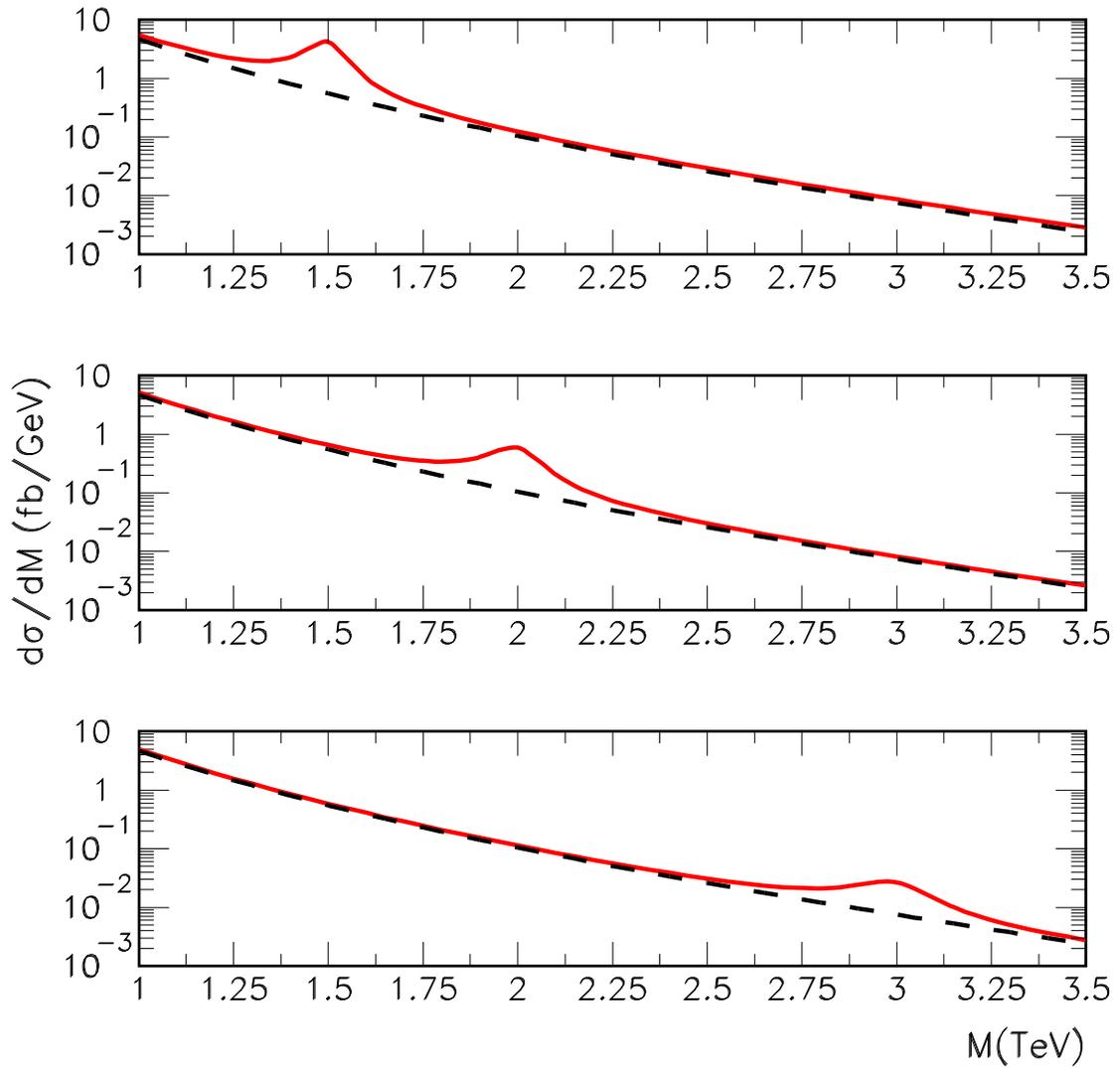}{0.99}
 \caption[Invariant mass distributions for single photon production at
 the LHC]{$d\sigma/dM$ (units of fb/GeV) {\em vs.} $M$ (TeV) is
   plotted for the case of the SM QCD background (dashed) and (first
   resonance) string signal + background (solid)~\cite{photons}.}
\label{fig:bump}
\end{figure}

\begin{figure}
 \postscript{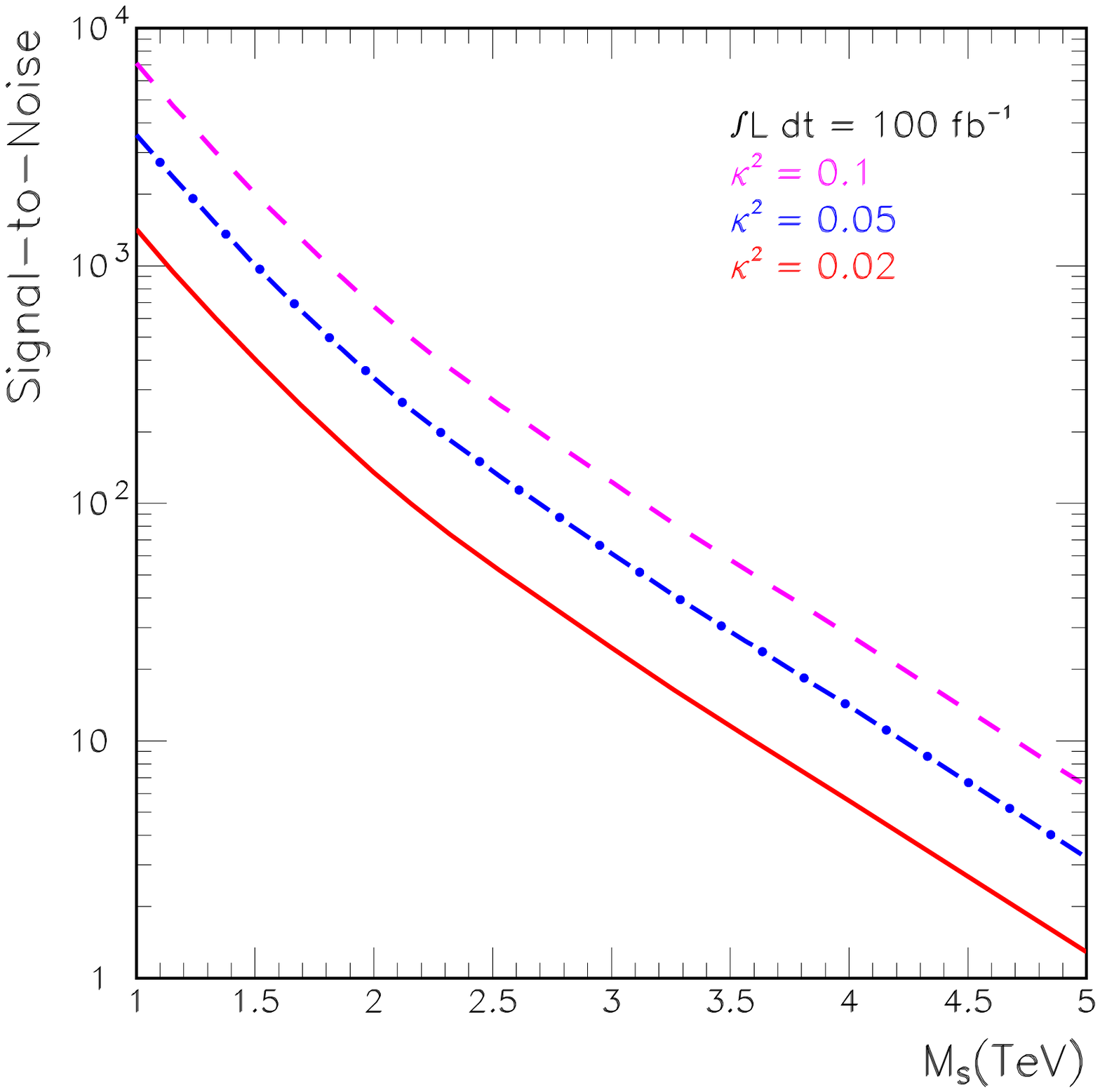}{0.98}
 \caption[$pp \to \gamma + {\rm jet}$ signal-to-noise
 ratio]{Signal-to-noise ratio for an integrated luminosity of
   100~fb$^{-1}$ and $\beta =2$.  The solid line is for $\kappa^2 =
   0.02$, the dot-dahsed line is for $\kappa^2 = 0.05$, and the dashed
   line is for an optimistic case with $\kappa^2 =
   0.1$~\cite{Anchordoqui:2008eu}.}
\label{fig:SN}
\end{figure}
In Fig.~\ref{fig:bump} we show several representative plots of this
cross section for different values of $M_s$. Standard bump-hunting
methods, such as calculating cumulative cross sections
\begin{equation}
\sigma (M_0) = \int_{M_0}^\infty  \frac{d\sigma}{dM} \, \, dM
\end{equation}
and searching for regions with significant deviations from the QCD
background, may allow for finding an interval of $M$ suspected of
containing a bump.  With the establishment of such a region, one may
calculate a signal-to-noise ratio, with the signal rate estimated in
the invariant mass window $[M_s - 2 \Gamma, \, M_s + 2 \Gamma]$. This
estimate of the signal would be roughly the same as that obtained
through the inclusive cut $k_\perp > 300$~GeV. This follows from the
relation~(\ref{relation}): for $M$ in the range of $M_s > 2$ and for
the significant contributing regions of $y$, the resulting $k_\perp$
cut in Eq.~(\ref{relation}) does not differ significantly from the estimated 
300~GeV. However, for $M_s > 2$~TeV, the background is significantly
reduced, augmenting the LHC discovery reach. In Fig.~\ref{fig:SN} we
show the signal-to-noise for different values of the mixing parameter
$\kappa$, assuming $\beta =2.$ It is clearly seen that even for
relatively small mixing, 100~fb$^{-1}$ of the LHC data could probe
deviations from the SM physics associated with TeV-scale strings at a
$5\sigma$ significance, for $M_s \lwig 4~{\rm TeV}$. Should bumps be
found, the D-brane model can be further differentiated from other
TeV-scale resonant processes by the details of the angular
distributions inherent in Eq.~(\ref{mhvlow2}).

\section{$Z$'s}

Analytic~\cite{Eardley:2002re} and numerical~\cite{Yoshino:2002tx}
studies have revealed that gravitational collapse takes place only at
sufficiently high energies and small impact parameters, as conjectured
years ago by Thorne~\cite{Thorne:ji}.  A horizon forms when and only
when a mass is compacted into a hoop whose circumference in every
direction is less than $2 \pi$ times its Schwarzschild radius up to a
factor of order 1.

The LHC black holes would decay largely via the Hawking 
process~\cite{Giddings:2001bu}, in which both  
the average number~\cite{Hawking:1974rv,Hawking:1974sw} and the
probability distribution of the
number~\cite{Parker:1975jm,Wald:1975kc,Hawking:1976ra} of outgoing
particles in each mode obey a thermal spectrum. In 10-dimensions, the
emission rate per degree of particle freedom $i$ of particles of spin
$s$ with initial total energy between $(\omega, \omega + d \omega)$ is
found to be~\cite{Han:2002yy}
\begin{equation}
\frac{\dot{N}_i}{d \omega} = \frac{\sigma_s (\omega)
\Omega_{d-3} \omega^{d-2}}{(d-2) (2\pi)^{d-1}} \left[
e^{\omega/T} - (-1)^{2s} \right]^{-1},
\label{rate}
\end{equation}
where $T = 7/(4\,\pi\,r)$ is the instantaneous Hawking temperature,
\begin{equation}
\Omega_{d-3} = \frac{2\,\pi^{(d-2)/2}}{\Gamma[(d-2)/2]}
\end{equation}
is the volume of a unit $(d-3)$-sphere, 
\begin{equation}
\label{schwarzins}
r = \frac{1}{M_{10}}
\left[ \frac{M}{M_{10}} \, 8 \, \pi^{3/2} \,\, \Gamma(9/2)
\right]^{1/7}
\end{equation}
is the instantaneous Schwarzschild radius of mass $M$~\cite{Myers:un},
and $\sigma_s (\omega)$ is the greybody absorption area due to the
backscattering of part of the outgoing radiation of frequency $\omega$
into the black hole (a.k.a. the greybody
factor)~\cite{Page:1976df}. The SM fields live on a 3-brane ($d=4$), while
gravitons inhabit the entire spacetime ($d=10$). The prevalent
energies of the decay quanta are of ${\cal O}(T \sim 1/r)$, resulting
in s-wave dominance of the final state. Indeed, as the total angular
momentum number of the emitted field increases, $\sigma_s (\omega)$
is rapidly 
suppressed~\cite{Kanti:2002nr,Ida:2002ez,Kanti:2002ge,Harris:2003eg}. In the low
energy limit, $\omega \, r \ll 1,$ higher-order terms are suppressed
by a factor of $3 (\omega\,r)^{-2}$ for fermions and by a factor of
$25 (\omega\,r)^{-2}$ for gauge bosons. For an average particle energy
$\langle \omega \rangle$ of ${\cal O}(r^{-1})$, higher partial waves
are also suppressed, although by a smaller factor. This strongly
suggests that the black hole is sensitive only to the radial coordinate and
does not make use of the extra angular modes available in the internal
space~\cite{Emparan:2000rs}. Actually, a recent detailed
analysis~\cite{Cardoso:2005vb,Cardoso:2005mh} has explicitly shown
that the relative emission rate of the SM particles and the 10-dimensional
bulk graviton is roughly 92:5. This implies that the power lost in the
bulk is less than 15\% of the total black hole mass, largely
favoring the dominance of visible decay. Therefore, in what follows, we
assume the Hawking evaporation process to be dominated by the SM brane
modes and we neglect graviton emission during the Schwarzschild phase.

Altogether, the average total emission rate for particle species $i$
is
\begin{equation}
\frac{d \av{N} }{dt}
= \frac{1}{2\pi} \left( \sum c_i\, g_i\, \Gamma_i \right)
\zeta(3)\, \Gamma(3)\, r^2\, T^3 \, ,
\label{N}
\end{equation}
where $c_i$ is the number of internal degrees of freedom of particle
species $i$, $g_i= 1\, (3/4)$ for bosons (fermions),
\begin{equation}
\Gamma_i = \frac{1}{4\pi r^2}
\int \frac{\sigma_s(\omega)\, \omega^2\, d\omega}
{e^{\omega/T}\pm 1} \left[ \int \frac{\omega^2\, d\omega}
{e^{\omega/T}\pm 1}\right]^{-1}  \, ,
\label{gamma}
\end{equation}
and $\Gamma_i = 0.60$ ($\Gamma_i = 0.66$) for bosons
(fermions)~\cite{Anchordoqui:2003ug}. This implies that black holes
decay with roughly equal probability to all degrees of freedom of the SM
particles. Since there are six charged leptons, one $Z$ boson, and one
photon, we expect $\sim 10\%$ of the particles to be hard primary
leptons and 2\% of the particles to be hard photons and $Z$'s, each
carrying hundreds of GeV of energy.

We now discuss some interesting contrast of $\gamma$ and $Z$
production in $D$-brane models that can serve as an additional marker
for discovery of string recurrences.  Ingnoring the $Z$-mass (i.e.,
keeping only transverse $Z$'s), and assuming that cross sections
$\times$ branching into lepton pairs are large enough for complete
reconstruction to $pp \to Z + {\rm jet},$ the quiver contribution to
the signal is suppressed relative to the photon signal by a factor of
$\tan^2\theta_W = 0.29.$ The SM ratio ($Z$ background)/( $\gamma$
background) is roughly 0.92 for processes involving $u$ (or $\bar u$)
quarks, and 4.7 for processes involving $d$ (or $\bar d$) quark.
Thus, even if $d$ quark processes are ignored, one obtains a
signal-to-noise ratio $ ({\rm S}/{\rm N})_Z=0.29/\sqrt{0.92}=0.30 \,
({\rm S}/{\rm N}_\gamma).$ Keeping the $d$ quarks will only lead to
more suppression of $({\rm S}/{\rm N})_Z$.\footnote{It is worth
  pausing to note that $\pi^0$ misidentification does not play a role
  in the $Z$ channel, and so this tends to decrease the QCD
  background. On the other hand, the string signal will suffer some
  suppression because of finite mass effects. These systematics (which
  have opposite effects on $(S/N)_Z$) were not considered in the
  preceding discussion.} This implies that if the high-$k_\perp$
photons, as predicted by the TeV string model, are discovered at
$5\sigma,$ they will not be accompanied by any significant deviation
of $pp \to Z + {\rm jet}$ from the SM predictions. This differs radically
from the evaporation of black holes produced at the LHC.  In that 
case, production of high-$k_\perp$ $Z$ and $\gamma$ are comparable.
The suppression of high-$k_\perp$ $Z$ production, whose origin lies in
the particular structure of the quiver model, will hold true for all
the low-lying levels of the string.

%% file: chap_dijets.tex
The string amplitudes that involve four matter fields depend on the
details of the D-brane geometry and how the D-branes are embedded into
the compact Calabi-Yau space. This is because modes of the internal
geometry can be exchanged during the four fermion scattering
processes. However, as shown in Ref.~\cite{stringhunter}, the poles of
the amplitudes for two gauge bosons and two matter fermions are due to
the exchanges of massless gauge bosons and universal string Regge
excitations only, and so computations of the respective average square
amplitudes can be performed in a model independent and universal way.
In this chapter, we extend our search for string signals at the LHC,
by including scattering processes which involve four gauge bosons as well as 
two gauge bosons and two fermions.

\section{$pp \to {\rm dijet}$}

As noted in Ref.~\cite{Meade:2007sz}, string signals are likely to
show up in the dijet channel.  The physical processes underlying dijet
production at the LHC are the collisions of two partons, producing two
final partons that fragment into hadronic jets. The corresponding
$2\to 2$ scattering amplitudes, computed at the leading order in
string perturbation theory, are collected in Ref.~\cite{stringhunter}.
The average square amplitudes are given by the following:
\begin{eqnarray}
|{\cal M} (gg \to gg)| ^2 & = & g^4 \left(\frac{1}{  s^2} + \frac{1}{  t^2} + \frac{1}{  u^2} \right) \left[ \frac{2 N^2}{N^2-1} \, (  s^2 \, V_s^2 +   t^2 \, V_t^2 +   u^2 \,  V_u^2) \right. \nonumber \\
 & + & \left. \frac{4 (3 - N^2)}{N^2 (N^2-1)} \, (  s \, V_s +   t \, V_t +   u \, V_u)^2 \right] \,,
\label{gggg}
\end{eqnarray}
\begin{equation}
|{\cal M} (gg \to q \bar q)|^2 =
g^4 N_f \frac{  t^2 +   u^2}{  s^2} \left[\frac{1}{2N} \frac{1}{  u \,   t} (  t \, V_t +   u \, V_u)^2 - \frac{N}{N^2 -1} \, V_t \, V_u \right] \,\,,
\label{ggqbarq}
\end{equation}
\begin{equation}
|{\cal M} (q \bar q \to gg)|^2 =
g^4 \ \frac{  t^2 +   u^2}{  s^2} \ \left[\frac{(N^2 -1)^2}{2 N^3} \ \frac{1}{  u   t} \ (  t \, V_t +   u \, V_u)^2 - \frac{N^2 -1}{N} \, V_t \, V_u \right] \, \,,
\label{qbarqgg}
\end{equation}
and 
\begin{equation}
|{\cal M}(qg \to qg)|^2 =
g^4 \ \frac{  s^2 +   u^2}{  t^2} \left[V_s \, V_u - \frac{N^2 -1}{2 N^2} \ \ \frac{1}{  s   u} \ (  s\, V_s +   u \, V_u)^2 \right] \,\,,
\label{qgqg}
\end{equation}
where the string ``formfactor'' functions of the Mandelstam variables are
defined as
\begin{equation}
V_t =V(  s,  t,  u) ~,\qquad V_u=V(  t,  u,  s) ~,\qquad  V_s=V(  u,   s,   t)~,
\end{equation}
with
\begin{equation}
V(  s,   t,   u)= {\Gamma(1-   s/M_s^2)\ \Gamma(1-   u/M_s^2)\over
    \Gamma(1+   t/M_s^2)}\ .
\end{equation}

Before proceeding, we pause to review our notation. The first Regge
excitations of the gluon $(g)$ and quarks $(q)$ will be denoted by
$G^*,\ q^*$, respectively. In the D-brane models under consideration,
the ordinary $SU(3)$ color gauge symmetry is extended to $U(3)$, so
that the open strings terminating on the stack of ``color'' branes
contain an additional $U(1)$ gauge boson $C^0$ and its excitations to
accompany the gluon and its excitations. The first excitation of the
$C^0$ will be denoted by $C^{0*}$.

In the following we isolate the contribution from the first resonant
state in Eqs.~(\ref{gggg}) - (\ref{qgqg}). For partonic center of mass
energies $\sqrt{s}<M_s$, contributions from the Veneziano functions
are strongly suppressed, as $\sim (\sqrt{s}/M_s)^8$, over standard
model processes; see Eq.~(\ref{mhvlow}). (Corrections to SM processes
at $\sqrt{s} \ll M_s$ are of order $(\sqrt{s}/M_s)^4$.) In order to
factorize amplitudes on the poles due to the lowest massive string states,
it is sufficient to consider $s=M_s^2$. In this limit, $V_s$ is
regular while
\begin{equation}
V_t=\frac{  u}{  s-M_s^2}~,\qquad V_u=\frac{  t}{  s-M_s^2}~.
\end{equation}
Thus the $s$-channel pole term of the average square amplitude
(\ref{gggg}) can be rewritten as\footnote{Note that the contributions 
  of single poles to the cross section are antisymmetric about the
  position of the resonance, and vanish in any integration over the
  resonance. Let the amplitude be $a +b/D$ in the vicinity of the
  pole, where $a$ and $b$ are real, $D = x+i\epsilon,$ $x=s-M_s^2,$
  and $\epsilon = \Gamma M_s.$ Then, since Re$(1/D) = x/|D|^2$, the
  cross section becomes $\sigma \propto a^2 + b^2/|D|^2 + 2 \, a \, b
  \, x/|D|^2 \simeq a^2 + b^2 \, \pi \, \delta(x)/\epsilon + 2ab \, \pi \, x
 \  \delta(x)/\epsilon$. Integrating over the width of the resonance,
  one obtains $a^2 \epsilon + b^2 \pi/\epsilon \simeq b \pi$,
  because $b \propto \epsilon$, $a \propto g^2$ and $\epsilon \propto
  g^2$.}
\begin{equation}
|{\cal M} (gg \to gg)| ^2  =  2 \ 
\frac{g^4}{M_s^4}\ \left(\frac{N^2-4+(12/N^2)}{N^2-1}\right) 
 \ \frac{M_s^8+  t^4 +   u^4}{(  s - M_s^2)^2} \, .
\label{ggggpole}
\end{equation}
The singularity at $s=M_s^2$ needs softening to a Breit-Wigner form,
reflecting the finite decay widths of resonances propagating in the
$s$ channel. Due to averaging over initial polarizations,
Eq.(\ref{ggggpole}) contains additive contributions from both spin
$J=0$ and spin $J=2$ $U(3)$ bosonic Regge recurrences ($G^*$ and
$C^{0*}$ in the notation of Ref.~\cite{widths}), created
by the incident gluons in the helicity configurations ($\pm \pm$) and
($\pm \mp$), respectively.  The $M_s^8$ term in Eq.~(\ref{ggggpole})
originates from $J=0$, and the $  t^4+   u^4$ piece reflects $J=2$
activity. Since the resonance widths depend on the spin and on the
identity of the intermediate state ($G^*$, $C^{0*}$) the pole term
(\ref{ggggpole}) should be smeared as
\begin{eqnarray}
\label{gggg2}
|{\cal M} (gg \to gg)| ^2 & = & 2\ 
\frac{g^4}{M_s^4}\ \left(\frac{N^2-4+(12/N^2)}{N^2-1}\right)  \\
 & \times & \left\{ W_{G^*}^{gg \to gg} \, \left[\frac{M_s^8}{(  s-M_s^2)^2 
+ (\Gamma_{G^*}^{J=0}\ M_s)^2} \right. \right.
\left. +\frac{  t^4+   u^4}{(  s-M_s^2)^2 + (\Gamma_{G^*}^{J=2}\ M_s)^2}\right] \nonumber \\
   & + &
W_{C^{O*}}^{gg \to gg} \, \left. \left[\frac{M_s^8}{(  s-M_s^2)^2 + (\Gamma_{C^{0*}}^{J=0}\ M_s)^2} \right.
\left. +\frac{  t^4+  u^4}{(  s-M_s^2)^2 + (\Gamma_{C^{0*}}^{J=2}\ M_s)^2}\right] \right\}, \nonumber
\end{eqnarray}
where $\Gamma_{G^*}^{J=0} = 75\, (M_s/{\rm TeV})~{\rm GeV}$,
$\Gamma_{C^{0*}}^{J=0} = 150 \, (M_s/{\rm TeV})~GeV$,
$\Gamma_{G^*}^{J=2} = 45 \, (M_s/{\rm TeV})~{\rm GeV}$, and
$\Gamma_{C^{0*}}^{J=2} = 75 \, (M_s/{\rm TeV})~{\rm GeV}$ are the
total decay widths for intermediate states $G^*$ and $C^{0*}$, with
angular momentum $J$. The associated weights of these two intermediate
states are given in terms of the probabilities for the various
entrance and exit channels
\begin{equation}
W_{G^*}^{gg \to gg} = \frac{(\Gamma_{G^* \to GG})^2}{(\Gamma_{G^* \to GG})^2 +
(\Gamma_{C^{0*} \to GG})^2} = 0.09
\label{w1}
\end{equation}
and 
\begin{equation}
W_{C^{0*}}^{gg \to gg} = \frac{(\Gamma_{C^{0*}
  \to GG})^2}{(\Gamma_{G^* \to GG})^2 + (\Gamma_{C^{0*} \to GG})^2} =
0.91 \, .
\label{w2}
\end{equation}
A similar calculation transforms Eq.~(\ref{ggqbarq}) near the pole into
\begin{eqnarray}
  |{\cal M} (gg \to q \bar q)|^2 & = & \frac{g^4}{M_s^4}\ N_f\ \left(\frac{N^2-2}{N(N^2-1)}\right)
  \left [W_{G^*}^{gg \to q \bar q}\, \frac{  u   t(   u^2+   t^2)}{(  s-M_s^2)^2 + (\Gamma_{G^*}^{J=2}\ M_s)^2} \right. \nonumber \\
  & + & \left. W_{C^{0*}}^{gg \to q \bar q}\, \frac{  u   t (   u^2+   t^2)}{(  s-M_s^2)^2 + 
      (\Gamma_{C^{0*}}^{J=2}\ M_s)^2} \right] \, ,
\end{eqnarray}
where 
\begin{equation}
W_{G^*}^{gg \to q \bar q}  = W_{G^*}^{q \bar q \to gg} = 
\frac{\Gamma_{G^* \to GG} \, 
\Gamma_{G^* \to q \bar q}} {\Gamma_{G^* \to GG} \, 
\Gamma_{G^* \to q \bar q} + \Gamma_{C^{0*} \to GG} \, 
\Gamma_{C^{0*} \to q \bar q}} = 0.24 
\label{w3}
\end{equation}
 and 
\begin{equation}
W_{C^{0*}}^{gg \to q \bar q} = W_{C^{0*}}^{q \bar q \to gg}  = 
\frac{\Gamma_{C^{0*} \to GG} \, 
\Gamma_{C^{0*} \to q \bar q}}{\Gamma_{G^* \to GG} \, 
\Gamma_{G^* \to q \bar q} + \Gamma_{C^{0*} \to GG} \, 
\Gamma_{C^{0*} \to q \bar q}} = 0.76 \, .
\label{w4}
\end{equation}
Near the $  s$ pole Eq.~(\ref{qbarqgg}) becomes
\begin{eqnarray}
|{\cal M} (q \bar q \to gg)|^2  & = &  \frac{g^4}{M_s^4}\ \left(\frac{(N^2 -2) 
(N^2-1)}{N^3}\right)
\left[ W_{G^*}^{q\bar q \to gg} \,  \frac{  u   t(   u^2+   t^2)}{(  s-M_s^2)^2 + (\Gamma_{G^*}^{J=2}\ M_s)^2} \right. \nonumber \\
 & + & \left.  W_{C^{0*}}^{q\bar q \to gg} \, \frac{  u   t(   u^2+   t^2)}{(  s-M_s^2)^2 + (\Gamma_{C^{0*}}^{J=2}\ M_s)^2} \right] \,\,,
\end{eqnarray}
whereas Eq.~(\ref{qgqg}) can be rewritten as
\begin{eqnarray}
|{\cal M}(qg \to qg)|^2 & = & - \frac{g^4}{M_s^2}\ 
\left(\frac{N^2-1}{2N^2}\right)
\left[ \frac{M_s^4   u}{(  s-M_s^2)^2 + (\Gamma_{q^*}^{J=1/2}\ M_s)^2}\right. \nonumber \\
 & + & \left. \frac{  u^3}{(s-M_s^2)^2 + (\Gamma_{q^*}^{J=3/2}\ M_s)^2}\right] \, \, .
\label{qgqg2}
\end{eqnarray}
The total decay widths for the $q^*$ excitation are: $\Gamma_{q^*}^{J=1/2} = \Gamma_{q^*}^{J=3/2} = 37\,
(M_s/{\rm TeV})~{\rm GeV}$~\cite{widths}.
Superscripts $J=2$ are understood to be inserted on all the $\Gamma$'s in
Eqs.~(\ref{w1}), (\ref{w2}), (\ref{w3}), (\ref{w4}).
Equation~(\ref{gggg2}) reflects the fact that weights for $J=0$ and
$J=2$ are the same~\cite{widths}. In what follows we set
$N=3$ and $N_f =6$.

\begin{figure}
 \postscript{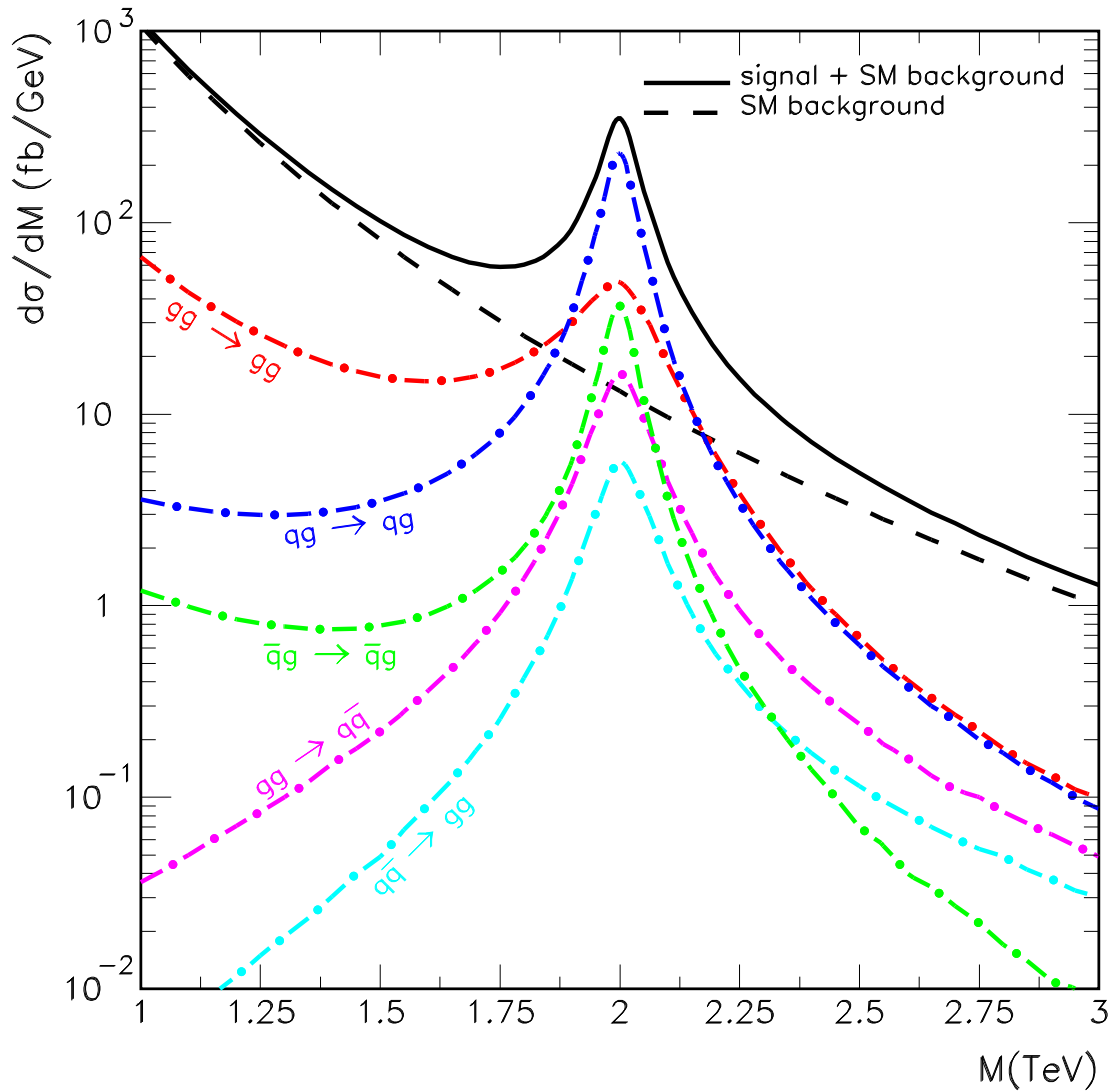}{0.98}
 \caption[$pp \to {\rm dijet}$ invariant mass spectrum]{$d\sigma/dM$
   (units of fb/GeV) {\em vs.}  $M$ (TeV) is plotted for the case of
   the SM QCD background (dashed line) and (first resonance) string signal
   + background (solid line). The dot-dashed lines indicate the
   different contributions to the string signal ($gg \to gg$, $gg \to
   q \bar q$, $qg \to qg$, and $q \bar q \to gg$)~\cite{dijets}.}
\label{fig:bump2}
\end{figure}

The resonance would be visible in data binned according to the
invariant mass $M$ of the dijet, after setting cuts on the different
jet rapidities, $|y_1|, \, |y_2| \le 1$~\cite{CMS} and transverse
momenta $p_{\rm T}^{1,2}>50$ GeV.  In Fig.~\ref{fig:bump2} we show a
representative plot of the invariant mass spectrum, for $M_s =2$~TeV,
detailing the contribution of each subprocess.  The QCD background has
been calculated at the partonic level from the same processes as
designated for the signal, with the addition of the $t$-channel
exchange process $qq\to qq$.  Our calculation, making use of the CTEQ6
parton distribution functions~\cite{Pumplin:2002vw,Stump:2003yu}
agrees with that presented in~\cite{CMS}.

We now estimate (at parton level) the LHC discovery reach. To do so,
we calculate a signal-to-noise ratio, with the signal rate estimated
in the invariant mass window $[M_s - 2 \Gamma, \, M_s + 2 \Gamma]$. As
usual, the noise is defined as the square root of the number of
background events in the same dijet mass interval for the same
integrated luminosity.

\begin{figure}
 \postscript{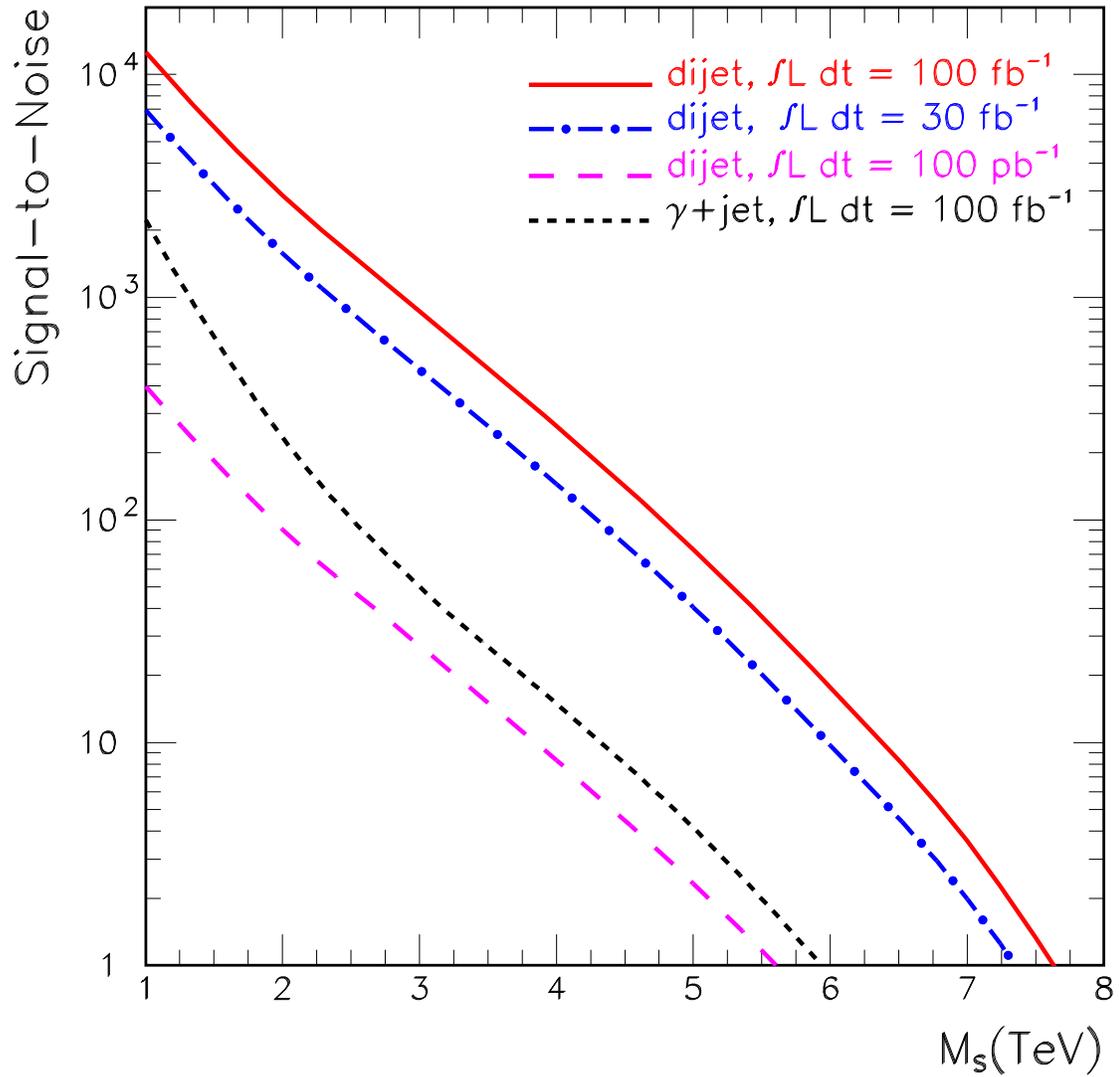}{0.98}
 \caption[$pp \to {\rm dijet}$ signal-to-noise ratio]{$pp \to {\rm
     dijet}$ signal-to-noise ratio for three integrated
   luminosities. For comparison, we also show the signal-to-noise of
   $pp \to \gamma + {\rm jet}$, for the minimal quiver stadard
   model~\cite{dijets}.}
\label{fig:S2N2}
\end{figure}

The top two and bottom curves in Fig.~\ref{fig:S2N2} show the behavior
of the signal-to-noise (S/N) ratio as a function of the string scale
for three integrated luminosities (100~fb$^{-1},$ 30~fb$^{-1}$ and
100~pb$^{-1}$) at the LHC. It is remarkable that within 1-2 years of
data collection, {\it string scales as large as 6.8 TeV are open to
  discovery at the $\geq 5\sigma$ level.}\footnote{This intersects
  with the range of string scales consistent with correct weak mixing
  angle found in the minimal quiver standard model~\cite{ant}.} For
30~fb$^{-1},$ the presence of a resonant state with mass as large as
5.7 TeV can provide a signal of convincing significance $(S/N >
13)$. The bottom curve, corresponding data collected in a very early
run of 100~$pb^{-1},$ shows that a resonant mass as large as 4.0~TeV
can be observed with $10\sigma$ significance! Once more, we stress
that these results contain no unknown parameters. They depend only on
the D-brane construct for the standard model, and {\it are independent
  of compactification details.}

The amplitudes for the four-fermion processes like quark-antiquark
scattering are more complicated because the respective formfactors
describe not only the exchanges of Regge states but also of heavy
Kaluza-Klein and winding states with a model-dependent spectrum
determined by the geometry of extra dimensions. Fortunately, they are
suppressed, for two reasons. First, the QCD $SU(3)$ color group
factors favor gluons over quarks in the initial state. Second, the
parton luminosities in proton-proton collisions at the LHC, at the
parton center of mass energies above~1 TeV, are significantly lower
for quark-antiquark subprocesses than for gluon-gluon and gluon-quark,
see Fig.~\ref{fig:pl}. The collisions of valence quarks occur at
higher luminosity; however, there are no Regge recurrences appearing
in the $s$-channel of quark-quark scattering~\cite{stringhunter}.

\section{$pp \to \gamma + {\rm jet}$}

In this section we estimate corrections from scattering of two
gauge bosons and two matter fermions to the $pp \to \gamma + {\rm
  jet}$ channel.  From our dijet analysis in the previous section and
Fig.~\ref{fig:pl}, it is easily seen that the average square amplitude
dominating $pp \to \gamma + {\rm jet}$ reads~\cite{stringhunter}
\begin{equation}
|{\cal M} (qg \to q \gamma) |^2 = 
- g^2 Q_c^2 \frac{1}{N} \frac{s^2 + u^2}{sut^2} \, 
(s\, V_s + u \, V_u)^2 \, ,
\end{equation}
where, as we defined in Chapter 3, $Q_c$ is the product of the $U(1)$
charge of the fundamental representation ($\sqrt{1/6}$) followed by
successive projections onto the hypercharge and then onto the photon
($\cos \theta_W$). For comparison with our dijet analysis, we also
show in Fig.~\ref{fig:S2N2} a fourth curve, for the process
$pp\rightarrow \gamma +$ jet, taking into account all the possible
contributions. The approximate equality of the background due to
misidentified $\pi^0$'s and the QCD background, across a range of
large $k_\perp$ as implemented in Chapter 4, is maintained as an
approximate equality over a range of invariant $\gamma$-jet invariant
masses with the rapidity cuts imposed. When considering contributions
from scattering processes with two gauge bosons and two fermions, the LHC
discovery reach in the $pp \to \gamma + {\rm jet}$ channel is extended
up to $M_s \sim 5.0$~TeV.

%% file: conclusion.tex
In the first part of this Dissertation,
we studied the six dimensional Salam-Sezgin model,
where a solution of the form Minkowski$_{4} \times S^2$ is known to
exist, with a $U(1)$ monopole serving as background in the two-sphere.
This model circumvents the hypotheses of the no-go
theorem of Maldacena and Nun\~ez, and then when lifted to string theory,
can show a dS phase. In our analysis we have allowed for time dependence
of the six-dimensional moduli fields and metric (with a
Robertson-Walker form).  Time dependence in these fields vitiates
invariance under the supersymmetry transformations.  With these
constructs, we have obtained the following results:

\begin{itemize}
\item In terms of linear combinations of the $S^2$ moduli field and
  the six dimensional dilaton, the effective potential consists of
  $(a)$ a pure exponential function of a quintessence field (this
  piece vanishes in the supersymmetric limit of the static theory) and
  $(b)$ a part which is a source of cold dark matter, with a mass
  proportional to an exponential function of the quintessence field.
  This presence of a VAMP CDM candidate is inherent in the model.
  
\item If the monopole strength is precisely at the value prescribed by
  supersymmetry, the model is in gross disagreement with present
  cosmological data - there is no accelerative phase, and the
  contribution of energy from the quintessence field is purely
  kinetic.  However, a miniscule deviation of ${\cal O}(10^{-120}$)
  from this value permits a qualitative match with data. Contribution
  from the VAMP component to the matter energy density can be as large
  as about 7\% without having negative impact on the fit.  The
  emergence of a VAMP CDM candidate as a necessary companion of dark
  energy has been a surprising aspect of the present findings, and
  perhaps encouraging for future exploration of candidates which can
  assume a more prominent role in the CDM sector.  

\item In our model, the exponential potential has behavior $V_Y\sim
  e^{\lambda Y/M_{\rm Pl}}$, with $Y$ the quintessence field and
  $\lambda = \sqrt{2}.$ The asymptotic behavior of the scale factor
  for exponential potentials is $e^{h(t)} \approx t^{2/\lambda^2},$ so
  that for our case $h \approx \ln t,$ leading to a conformally flat
  Robertson-Walker metric for large times.  The evolution from
  constant velocity expansion to a brief accelerated phase in the
  neighborhood of our era makes the model phenomenologically viable.
  In the case that the supersymmetry condition ($b^2\xi =1$) is
  imposed, and there is neither radiant energy nor dark matter except
  for the $X$ contribution, we find for large times that the growth of
  the scale factor is given by $e^{h(t)} \approx \sqrt{t},$ so that
  even in this case the asymptotic metric is Robertson-Walker rather
  than Minkowski.  Moreover, and rather intriguingly, the scale
  factor is what one would find with radiation alone.

\end{itemize}

In sum, in spite of the shortcomings of the model (not a perfect fit,
requirement of a tiny deviation from supersymmetric prescription for
the monopole embedding), it has provided  stimulating new, and
unifying, look at the dark energy and dark
matter puzzles.\\

On a separate track, the LHC program will include the identification
of events with single prompt high-$k_\perp$ photons as
probes of new physics. In the second part of this Thesis, we have shown that
this channel is uniquely suited to search for experimental evidence
of TeV-scale open string theory.  At the parton level, we analyzed
single photon production in gluon fusion, $gg \to \gamma g$, with open
string states propagating in intermediate channels. If the photon
mixes with the gauge boson of the baryon number, which is a common
feature of D-brane quivers, the amplitude appears already at the
string disk level. It is completely determined by the mixing parameter
(which is actually determined in the minimal theory) -- and it is
otherwise model-(compactification-) independent. We calculated cross
sections for Regge recurrences of fundamental strings, as well as the
QCD background.  (A vital part of the background discussion concerned
the minimization of misidentified $\pi^0$'s emerging from
high-$p_\perp$ jets.) We showed that even for relatively small mixing,
100~fb$^{-1}$ of the LHC data could probe deviations from the SM
physics associated with TeV-scale strings at a $5\sigma$ significance,
for $M_{\rm string}$ as large as 4~TeV. 

Another channel that can provide a clean signal of new physics at the
LHC is $pp \to {\rm dijet}$.  In D-brane constructions, the dominant
contributions to full-fledged string amplitudes for all the common QCD
parton subprocesses leading to dijets are completely independent of
the details of compactification, and can be evaluated in a
parameter-free manner. We made use of these amplitudes evaluated near
the first resonant pole to determine the discovery potential of the LHC
for the first Regge excitations of the quark and gluon. We found that,
remarkably, the reach of the LHC after a few years of running can be as
high as 6.8~TeV. Even after the first 100~pb$^{-1}$ of integrated
luminosity, string scales as high as 4.0~TeV can be discovered. For
string scales as high as 5.0~TeV, observations of resonant structures
in $pp\rightarrow \gamma~ +$ jet (considering parton subprocesses $gg \to \gamma g,$ $g q \to \gamma q,$
$g \bar q  \to \gamma \bar q $, and $q \bar q \to \gamma g$),
can provide interesting corroboration
of string physics at the TeV-scale. \\

All in all, a new era, with experimental measurements of string physics,
may be close at hand.